%% file: main.tex
\documentclass{fundam}

\publyear{22}
\papernumber{2102}
\volume{185}
\issue{1}

\usepackage{url} 
\usepackage[ruled,lined]{algorithm2e}
\usepackage{graphicx}
\usepackage{tikz}
\usetikzlibrary{automata, positioning, arrows, arrows.meta}

\usepackage{hyperref} 

\usepackage{amsmath}
\usepackage{amssymb}
\usepackage{graphicx}
\usepackage{theoremref}
\usepackage{enumitem}
\usepackage{moreenum}
\usepackage{mathrsfs}
\usepackage{mathtools}
\usepackage{thmtools}
\usepackage{adjustbox}
\usepackage{subcaption}
\usepackage{standalone}

\usepackage{tikz-cd}
\usepackage{pgfplots}
\pgfplotsset{compat=1.12}

\usepackage{macros}

\begin{document}

\title{Probabilistic automatic complexity of finite strings}
\runninghead{K.\ Gill}{Probabilistic automatic complexity}

\address{\url{gillmathpsu@posteo.net}}

\author{Kenneth Gill
\thanks{Most of this work is part of the author's Ph.D.\ dissertation at Penn State
University \cite{mythesis}. 
The author is grateful to his thesis advisors Jan Reimann and Linda
Westrick for their invaluable help and support, to Jake Canel for
several helpful conversations, and to Bj{\o}rn Kjos-Hanssen for suggesting the
proof of \thref{tarski} as well as inspiring the project as a whole.
This research was supported in part by NSF grant DMS-1854107.
}}
\date{June 3, 2024}

\maketitle

\begin{abstract}
  We introduce a new complexity measure for finite strings using probabilistic
  finite-state automata (PFAs), in the same spirit as existing notions employing
  DFAs and NFAs, and explore its properties. 
  The PFA complexity $A_P(x)$ is the least number of states of a PFA for which
  $x$ is the most likely string of its length to be accepted.
  The variant $A_{P,\delta}(x)$ adds a real-valued parameter $\delta$ specifying
  a required lower bound on the gap in acceptance probabilities between $x$ and
  other strings.
  We prove $A_{P,\delta}$ is $\delta$-computable for all $\delta$, relate $A_P$
  to the DFA and NFA complexities, and obtain a complete
  classification of binary strings with $A_P=2$.
  Finally, we discuss several other variations on $A_P$ with a view to
  obtaining additional desirable properties.
\end{abstract}


\input{intro}
\input{def}
\input{firstresults}
\input{classifs}
\input{classforward}
\input{classreverse}
\input{classafter}
\input{computability}
\input{otherdefs}

\bibliographystyle{fundam}
\bibliography{PFAsources}

\end{document}

%% file: intro.tex
\section{Introduction}

Informally, the Kolmogorov complexity of a finite string $w$ is the size of the
smallest Turing machine which outputs $w$ given no input. As a function, it is
well-known to be noncomputable, and moreover only defined up to an additive
constant. 
These drawbacks have motivated several authors to define complexity measures
based on models of computation less powerful than the Turing machine, such as
context-free grammars \cite{Diwan,smallest}.
In 2001, Shallit and Wang introduced one such measure using deterministic
finite-state automata (DFAs), defining $A_D(w)$ to be the number of states of
the smallest DFA for which $w$ is the only string of its length to be accepted
\cite{SW01}. This measure is computable, well-defined, and there is a polynomial-time
algorithm to recover $w$ from a witness for $A_D(w)$. Later in 2013, Hyde
defined a similar measure replacing DFAs with nondeterministic
finite-state automata (NFAs) \cite{H13,HK15}. $A_N$ shares the advantages of $A_D$
over Kolmogorov complexity while additionally making $A_N(w) = A_N(\rev{w})$,
where $\rev{w}$ is the reversal of $w$, and avoiding ``dead states''
(nonaccepting states with no out-transitions) often present among witnesses for
$A_D$ merely to satisfy the requirement of determinism. The study of $A_N$ has
been continued by Kjos-Hanssen, see e.g.~\cite{KIncompr,K20, KMaximal,KComplexity},
as well as the recent book \cite{KBook}.

Inspired by the aforementioned work, we investigate what happens when 
deterministic or nondeterministic machines are replaced by probabilistic ones
(PFAs), wherein each state transition occurs with some probability and each word
$w$ is assigned a probability of acceptance $\rho_M(w)$ by the PFA $M$. We view
$M$ as describing $w$ if $w$ uniquely maximizes $\rho_M$ among all strings
of the same length. This property can be phrased in terms of the so-called gap
function, which measures how well $M$ separates $w$ from other strings:
\begin{defn}
  The \bdef{gap function} of $M$ is the map from $\Sigma^\ast$ to $[-1,1]$ given by
  \[
    \gap_M(w) = \min \set{ \rho_M(w) - \rho_M(z) : \abs{z} = \abs{w} \text{ and
    } z\neq w}.
  \]
\end{defn}

All words are presumed to be drawn from some finite alphabet $\Sigma$ fixed
in advance. 
The gap function is positive iff $w$ is more likely than any other string of the
same length to be accepted, and we define the PFA complexity to be the least
number of states needed for this to happen:
\begin{defn}\thlabel{apdef}
  The \bdef{probabilistic automatic complexity} (PFA complexity) of $w$ is
  \[
    A_P(w) = \min\set{k \st \text{there is a $k$-state PFA $M$ such that
    $\gap_M(w) > 0$}}.
  \]
\end{defn}

This definition is probably the one most directly analogous to the definitions
of $A_D$ and $A_N$, and it turns out that $A_P$ is also computable, as $A_D$ and
$A_N$ are (\thref{apcomp}). 
One peculiarity of $A_P$, however, is that $M$ can witness $A_P(w)$
while $\rho_M(w)$ is not very high, or not much different from $\rho_M(z)$ for
other strings $z$ of the same length. Is $M$ really a good representation of $w$
if it can only slightly distinguish $w$ from other strings? What if all
potential witnesses $M$ have this property? 
We address this problem by introducing a real-valued parameter giving a lower
bound on the gap between probabilities:

\begin{defn}\thlabel{apdeltadef} 
  The \bdef{probabilistic automatic complexity of $w$ with gap}
  $\delta \in [0,1)$ is
  \[
    A_{P,\delta}(w) = \min\set{ k \st \text{there is a $k$-state PFA $M$ such that
    $\gap_M(w) > \delta$}}.\footnote{This is a slightly different definition from
    that originally given by the author in \cite{mythesis}, which required
    $\gap_M(w)\geq \delta$ rather than $>$. The author has come to feel that the
    present definition is more natural. Only minor amendments to the proofs of
    results involving $A_{P,\delta}$ were needed as a result of this change.}
  \]
\end{defn}

Thus $A_{P,0}(w) = A_P(w)$. \thref{apdeltacomp,tarski} together imply that
$A_{P,\delta}$ is $\delta$-computable for all $\delta$. 
We in fact establish the almost-everywhere uniform computability of
$A_{P,\delta}(w)$ as a function of two variables in \thref{apdeltacomp},
although the argument does not extend to the case $\delta=0$.

Our other main result about $A_P$ is the following complete classification of
binary strings with complexity 2, which arguably helps to vindicate $A_P$ by
showing that it does appear to capture some intuitive structure in strings:

\begin{restatable*}{rethm}{classtwo}\thlabel{classification2-full}
  For $w \in \{i,j\}^\ast$, we have $A_P(w)=2$ if and only if $w$ is of the form
  \[
    i^nj^m, \qquad i^nj^mi, \qquad i^n(ji)^m, \qquad\text{or}\quad i^nj(ij)^m
  \]
  for some $n,m\geq0$.
\end{restatable*}

One can compare this to the fact that $A_N(w) = 2$ if and only if $w = ij^n$,
$i^nj$, or $(ij)^n$ \cite[p.~18]{H13}. Indeed, $A_N$ is unbounded on strings of the
form $i^nj^m$, which implies

\begin{restatable*}{recor}{ANAPdiff}\thlabel{ANAPcor}
  The quantity $A_N(w) - A_P(w)$ may be arbitrarily large among binary $w$.
\end{restatable*}

$A_{P,\delta}$ has a philosophically attractive feature not shared by
$A_P$ which we now describe. 
Suppose one is given an automaton $M$ as a ``black box'', that is,
with no information whatsoever about its inner workings. All one can do is run
it with some input string, and check whether it accepts or rejects the string.
Suppose further that an experimenter wishes to test whether this automaton
witnesses an upper bound for $A_P(w)$ for some string $w$.
Then the experimenter needs not only to check whether each $z \in
\Sigma^{\abs{w}}$ is accepted, but whether or not it will be accepted with a
lower bound $\lambda$ on its probability of acceptance, for each $\lambda$ in
turn. (This would enable them to decide if there is some particular $w,\lambda$
with $\rho_M(w) > \lambda$ but $\rho_M(z) < \lambda$ whenever $\abs{z}=\abs{w}$
and $z\neq w$. In other words, they would estimate a lower bound on $\gap_M(w)$.)
The experimenter can only accomplish this by running the machine repeatedly on
each input $w$ to get some sense of the expected value of $\rho_M(w)$, up to
some acceptable margin of error $\varepsilon$. 

In his original paper introducing PFAs, Rabin \cite{R63} discusses a similar
endeavor in the context of establishing experimentally that $w$ is in a given
stochastic language, where a language is stochastic if it is of the form $\set{
w\in \Sigma^\ast \st \rho_M(w) > \lambda}$ for some PFA $M$ and $\lambda \in
[0,1]$, called the \bdef{cut-point}. As he points out, the law of large numbers
implies that as long as $\rho_M(w) \neq \lambda$, there is a finite number
$N=N(w, \varepsilon)$ such that running $N$ trials, counting the number $s$ of
successes, and comparing $s/N$ with $\lambda$ will correctly determine if
$\rho_M(w)>\lambda$ with probability $1-\varepsilon$. But, as he goes on to say,
finding $N(w,\varepsilon)$ would depend on knowing $\rho_M(w)$ in the first
place.

Rabin's solution is to only consider cut-points $\lambda$ which are isolated for
$M$, that is, such that $\abs{ \rho_M(w) - \lambda}\geq \delta$ for all
$w\in\Sigma^\ast$ and some $\delta>0$. If one wants to run the above experiment
to test if $\rho_M(w)>\lambda$ when $\lambda$ is isolated, then the number of
trials $N$ needed to determine this within margin of error $\varepsilon$ now
only depends on $\delta$ and $\varepsilon$, regardless of $M$. Knowledge of
$\rho_M(w)$ is not needed. Of course, this is not a solution from a
practical point of view if no such cut-point is given at the outset, because now
the experimenter would need to determine if $\lambda$ is isolated for $M$ and
(if so) a lower bound for its degree of isolation $\delta$. The problem of
determining if a given rational cut-point is isolated for a given PFA is known
to be $\Sigma^0_2$-complete \cite[Theorem~1]{CSV13}. 

But---back to our black-box experiment---if one specifies $\delta$ at the outset
and looks for a witness for an upper bound on $A_{P,\delta}(w)$ rather than
$A_P(w)$, the problem disappears and we still get that $N$ depends only on
$\delta$ and $\varepsilon$, with both of these parameters now being chosen by
the experimenter. To see why, let a single trial consist of running every word
of length $\abs{w}$ through the machine $M$ once. If $s(w,N)$ is the number of
acceptances of $w$ in $N$ trials, then there is a function
$N=N(\delta,\varepsilon)$ such that for each $z\in \Sigma^{\abs{w}}$, one
correctly concludes with probability at least $1-\varepsilon'$ that
$\rho_M(w)-\rho_M(z) > \delta$ given $[s(w,N)-s(z,N)]/N > \delta$, assuming
$\rho_M(w) - \rho_M(z) \neq \delta$. Here $\varepsilon'$ is chosen small enough
that $(1 - \varepsilon')^{ \abs{\Sigma}^{ \abs{w} } } > 1- \varepsilon$. Since
acceptances and rejections of words are presumed to be independent events, it
follows that after $N(\delta,\varepsilon)$ trials, one correctly concludes with
probability at least $1-\varepsilon$ that $\gap_M(w) > \delta$.

This paper establishes several basic properties of $A_P$ and $A_{P,\delta}$ and
hopefully justifies their study as having intrinsic interest, but many
avenues of investigation are left unexplored. 
This is in part due to both $A_P$ and $A_{P,\delta}$ proving somewhat
combinatorially difficult to reason with, aside from a few of our results which
follow quickly from straightforward matrix calculations. In particular there is
nothing we can say about the asymptotic behavior of either quantity: no example
is known at the time of writing which even suggests $A_P$ can be greater than
$3$. Indeed, the original motivation behind proving the classification theorem
(\thref{classification2-full}) was to show that $A_P$ can be greater than $2$,
which was unclear up to that point.
Then the most fundamental question we leave unanswered is probably
\begin{restatable}{que}{unbndq} Is $A_P$ unbounded? If not, what is its maximum
  value? Similarly when restricted to a given finite alphabet, and similarly for
  $A_{P,\delta}$.
\end{restatable}

The structure of the rest of the paper is as follows. After collecting
some formal definitions in the next section, we state a few preliminary results
in Section~\ref{sec:firstresults}, including
\thref{upperbound} relating $A_P$ and $A_{P,\delta}$ to $A_D$ and $A_N$.
Here we also discuss a few examples of the calculation of $A_P$ and
$A_{P,\delta}$.
Section~\ref{sec:class}, which takes up over half of the paper, consists
entirely of the proof of \thref{classification2-full}. This proof exploits a
correspondence between PFAs and iterated function systems outlined in
Section~\ref{sec:ifs}.
Section~\ref{sec:apcomp} is devoted to proving the computability of $A_P$
and $A_{P,\delta}$, which involves techniques from computable analysis as well
as an application of a classical result in model theory.
Finally, in Section~\ref{sec:otherdefs} we discuss several further variations on
$A_P$ with an eye to mitigating its potential flaws as a complexity measure.

%% file: def.tex
\section{Preliminaries}\label{sec:defs}
Our notation is mostly standard.
Let $\Sigma^\ast$ be the set of finite strings over the finite alphabet
$\Sigma$. 
Write either $xy$ or $x^\smallfrown y$ for the concatenation of the strings $x$
and $y$.

\begin{defn} A \bdef{probabilistic finite-state automaton} (PFA) is an abstract
  machine specified by a tuple $M= (S, \Sigma, P, \vec{\pi}, \vec{\eta})$, where
  \begin{itemize}
    \item $S=\set{s_1,\dotsc,s_n}$ is the set of states;
    \item $\Sigma$ is a finite alphabet;
    \item $P$ is a set of $n\times n$ right-stochastic matrices $\set{P_\sigma
      \stc \sigma \in\Sigma}$ describing the transition probabilities. For each
      $\sigma \in\Sigma$, $(P_\sigma)_{ij}$ is the probability of going from $s_i$
      to $s_j$ when letter $\sigma$ is read;
    \item $\vec{\pi}$ is a row vector of length $n$ giving a probability
      distribution on initial states, so $\vec{\pi}_j$ is the probability of the
      machine starting in state $s_j$; and
    \item $\vec{\eta}$ is a column vector of length $n$ determining the set of
      accepting states, with $\vec{\eta}_i$ being $1$ if $s_i$ is accepting and
      $0$ otherwise.
  \end{itemize}
\end{defn}

\begin{figure}
  \centering
  \begin{tikzpicture}
    \node (s1) [state] {$s_1$};
    \draw[<-] (s1) -- node[above] {(1)} ++(-1.5cm,0);
    \node (s2) [state] at (3,-1.5) {$s_2$};
    \node (s3) [state,accepting] at (6,0) {$s_3$};
    \path [-stealth] (s1) edge[bend angle=20,bend left] node[above] {0 (1); 1
      (.5)} (s3) 
      (s1) edge[loop below] node {1 (.5)} () 
      (s2) edge[loop below] node {1 (1)} () 
      (s2) edge node[below,sloped] {0 (1)} (s1) 
      (s3) edge node[below=0mm] {0 (1); 1 (.5)} (s1) 
      (s3) edge node[right=1mm,below,sloped] {1 (.5)} (s2); 
  \end{tikzpicture}
  \caption[An example of a PFA.]{An example of a PFA. Numbers in parentheses are
    transition probabilities, so that the PFA starts in state $s_1$ with
    probability 1. $s_3$ is the unique accepting state.}
  \label{fig:pfaex} 
\end{figure}
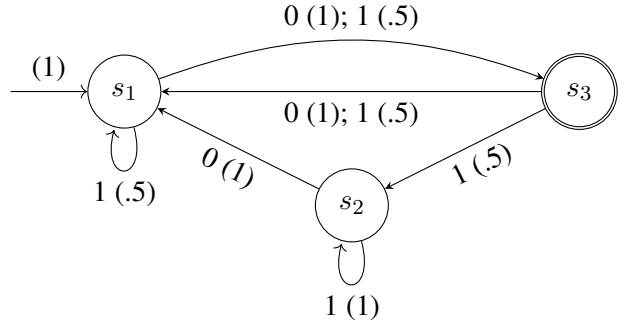

When $\Sigma$ is not important, we will frequently omit its mention, and
likewise we usually identify $S$ with the set $\{1,\dotsc,n\}$ for some $n$.
Thus we often specify a PFA by giving only $\vec{\pi}$, $\vec{\eta}$, and the
matrices $P_\sigma$.
If all entries of $\vec{\pi}$ and each $P_\sigma$ are rational numbers, then we
refer to $M$ as \bdef{rational}. A PFA can also be represented as a digraph, with
edges labeled by transition probabilities. For example, Figure~\ref{fig:pfaex}
depicts the PFA over the alphabet $\{0,1\}$ with
\[
    \quad P_0 =
    \begin{pmatrix} 0 & 0 & 1\\
      1 & 0 & 0\\
      1 & 0 & 0
    \end{pmatrix}, \quad P_1 = \begin{pmatrix} .5 & 0 & .5\\
      0 & 1 & 0\\
      .5 & .5 & 0
    \end{pmatrix}, \quad
    \vec{\pi} = (1,0,0), \quad \vec{\eta} = \begin{pmatrix} 0\\0\\1
    \end{pmatrix}.
\]

\begin{defn} If $M$ is a PFA and $x=x_1x_2\cdots x_\ell$ is a string, let
  \[
    P_M(x) = P_{x_1}P_{x_2}\cdots P_{x_\ell}.
  \]
  Then the \bdef{acceptance probability of $x$} with respect to $M$ is
  \[
    \rho_M(x) = \vec{\pi} P_M(x) \vec{\eta}.
  \]
\end{defn}

If $M$ is understood from context we may simply write $\rho(x)$, and similarly
$\gap(x)$.

One can view a DFA as the special case of a PFA in which $\vec{\pi}$ is a
coordinate vector and all $P_\sigma$s are permutation matrices. An NFA is then a
slight relaxation of a DFA where $\vec{\pi}$ may be any zero-one vector and each
$P_\sigma$ may be any zero-one matrix. Of course, DFAs and NFAs are usually
represented as digraphs, but it is convenient for us to think of them via their
transition matrices since we will manipulate them directly alongside PFAs. The
precise definitions of the DFA and NFA complexities are as follows:

\begin{defn}[Shallit and Wang \cite{SW01}] The \bdef{deterministic automatic
  complexity} of a finite string $x$ is
  \begin{align*}
    A_D(x) = \min \{k :~ &\text{there is a $k$-state DFA accepting $x$}\\
    & \text{uniquely among strings of length $\abs{x}$}\}.
  \end{align*}
\end{defn}
In other words, thinking of a witnessing DFA $M$ as a PFA, this says $\gap_M(x)
= 1$, or equivalently $\gap_M(x) > 0$ since the gap function takes only the
values 0 and 1 when $M$ is deterministic. It follows immediately that $A_P(x) \leq
A_D(x)$ for all $x$.

\begin{defn}[Hyde \cite{H13}] The \bdef{nondeterministic automatic complexity}
  of $x$ is
  \begin{align*}
    A_N(x) = \min\{k :~&\text{there is a $k$-state NFA accepting $x$}\\
           &\text{and with a unique accepting path of length $\abs{x}$}\}.
  \end{align*}
\end{defn}

Every DFA witnessing $A_D(x)$ is an NFA that accepts $x$ with a unique accepting
path of length $\abs{x}$ (by virtue of its determinism), so $A_N(x)\leq A_D(x)$
for all $x$.

\begin{rmk}
Probabilistic finite automata were independently introduced in 1963 by Michael
Rabin and J.~W.~Carlyle \cite{R63,Car63}. 
Carlyle's stochastic sequential machines are transducers with
both input and output behavior, while Rabin's PFAs---which are sometimes also 
called stochastic acceptors---can only accept an input string with some
probability. The present work focuses only on PFAs as defined by Rabin,
although Carlyle-style machines have found wide applicability in machine
learning and pattern recognition; see \cite{Vidal} for a modern survey. A notion
of transducer complexity of finite strings has also been studied \cite{CSR}, but
the approach taken there is most like that of the Kolmogorov complexity rather
than $A_D$. We leave the probabilistic analogue for future work.
There is, however, an idea related to $A_P$ which has been studied for
transducers in the machine learning literature.
Given a probabilistic finite-state transducer $T$, $x$ is called the \bdef{most
probable string} or \bdef{consensus string} of $T$ if it is generated by $T$
with maximal probability among all strings, not just among those with the same
length \cite{Vidal}. One might ask if this notion should be adapted to PFAs,
defining the complexity of $x$ instead as the smallest size (in some sense) of a
PFA accepting $x$ with unique highest probability among all strings. But we will
see in the proof of \thref{classification2} (Section~\ref{sec:classforward})
that a single PFA can simultaneously witness $A_P(x)$ for every one of an
infinite family of strings $x$ of similar structure. This ability arguably lends
$A_P$ a descriptive advantage over a notion resulting from viewing a PFA as only
describing its most probable string.
\end{rmk}

%% file: firstresults.tex
\section{First results on $A_P$}\label{sec:firstresults}

In this section we establish a few basic properties of $A_P$ and $A_{P,\delta}$,
beginning by relating them to $A_D$ and $A_N$:
\begin{prop}\thlabel{upperbound} \begin{enumerate}[label=(\roman*),leftmargin=*]
  \item For any $x$, $A_P(x)\leq A_N(x)+1$.
  \item $A_P(x) \leq A_{P,\delta}(x) \leq A_D(x)$ for all $x$ and $\delta \in [0,1)$.
    For every $x$, there is a $\delta'>0$ such that $A_{P,\delta}(x)=A_P(x)$ for
    all $\delta\in [0,\delta')$.
  \end{enumerate}
\end{prop}
\begin{proof} \begin{enumerate}[label=(\roman*),leftmargin=*]
  \item Let $M=(S,\Sigma,P,\vec{\pi},\vec{\eta})$ be an NFA witnessing $A_N(x)$.
    Uniqueness of $M$'s accepting path for $x$ means in particular that
    $\vec{\pi}$ is a coordinate vector.
    Then define a PFA $M' = (S',\Sigma, P',\vec{\pi}', \vec{\eta}^{\,'})$ as
    follows. Let $S'=S\cup \{s\}$, where $s$ is a new state not occurring in
    $S$, to be listed after all other states. Let $\vec{\pi}'= [ \vec{\pi} | 0]$
    and $\vec{\eta}^{\,'}=[\vec{\eta} | 0]$, where $[\vec{a} | \vec{b}]$ denotes
    the concatenation of the vectors $\vec{a}$ and $\vec{b}$.
    Write $X^i$ for the $i$th row of any matrix $X$ ($i\geq 1$). 
    For each $\sigma \in\Sigma$, let $P_\sigma'$ be built as follows from
    $P_\sigma$: if $P_\sigma^i$ has at least one nonzero entry, let
    $(P_\sigma')^i = [P_\sigma^i | 0] / (\sum P_\sigma^i)$. Otherwise, let
    $(P_\sigma')^i = [P_\sigma^i | 1] = (0, \dotsc, 0, 1)$. Finally, if $\abs{S}
    = k$, then append a new row $(P_\sigma')^{k+1} = (0,\dotsc, 0,1)$ (this
    corresponds to the new state $s$).

    Then $M'$ still has a unique accepting path of length $\abs{x}$; in
    particular, $x$ is the only string of length $\abs{x}$ with $\rho_{M'}(x)$
    positive. Therefore $M'$ witnesses an upper bound for $A_P(x)$.

  \item If $\delta<\delta'$ then $A_{P,\delta}(x) \leq A_{P,\delta'}(x)$, and if
    $M$ is a DFA witnessing $A_D(x)$ then $\gap_M(x) = 1$. This gives the
    first statement. For the second statement, one can for example pick $\delta'
    = \gap_M(x)/2$ for any witness $M$ for $A_P(x)$.
  \end{enumerate}
\end{proof}

\begin{cor} For all $x$, $A_P(x) \leq \floor{\abs{x}/2}+2$.
\end{cor}
\begin{proof} Hyde showed in \cite[Theorem~3.1]{H13} that $A_N(x) \leq
  \floor{\abs{x}/2} + 1$, so the bound immediately follows from the proposition.
\end{proof}

The procedure described in the first part of \thref{upperbound} demonstrates
that if $A_N(x)$ is witnessed by an NFA such that every state has at least one
out-transition for every letter, then $A_P(x)\leq A_N(x)$ (because there are no
rows of all zeros in the transition matrices, and the ``dead state'' $s$ need
not be added).

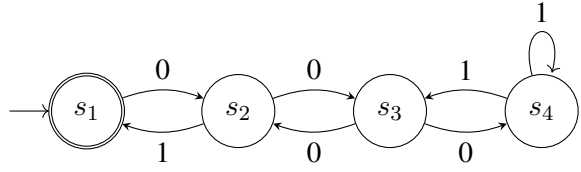
\begin{figure}
  \centering
    \begin{tikzpicture}
    \node (s1) [state,initial, accepting,initial text=] {$s_1$};
    \node (s2) [state] at (2,0) {$s_2$};
    \node (s3) [state] at (4,0) {$s_3$};
    \node (s4) [state] at (6,0) {$s_4$};
    \path [-stealth] (s1) edge[bend angle=20,bend left] node[above] {0} (s2)
    (s2) edge[bend angle=20,bend left] node[above] {0} (s3)
    (s3) edge[bend angle=20,bend left] node[below] {0} (s2)
      (s3) edge[bend angle=20,bend right] node[below] {0} (s4)
      (s2) edge[bend angle=20,bend left] node[below] {1} (s1)
      (s4) edge[bend angle=20,bend right] node[above] {1} (s3)
      (s4) edge[loop above] node {1} ();
  \end{tikzpicture}
  \caption{An NFA witnessing that $A_N(0001101)=4$.}
\label{fig:thenfa}
\end{figure}

As an example of this construction, according to Bj{\o}rn Kjos-Hanssen's
website,\footnote{\url{https://math.hawaii.edu/wordpress/bjoern/complexity-of-0001101/}}
$A_N(0001101)=4$ via the NFA depicted in Figure~\ref{fig:thenfa}. Here $s_1$ is
both the initial and accepting state. In matrix form, this NFA can be
represented as
\begin{equation}
  P_0 = \begin{pmatrix} 0 & 1 & 0 & 0\\
    0 & 0 & 1 & 0\\
    0 & 1 & 0 & 1\\
    0 & 0 & 0 & 0
  \end{pmatrix}, \quad
  P_1 = \begin{pmatrix} 0 & 0 & 0 & 0\\
      1 & 0 & 0 & 0\\
      0 & 0 & 0 & 0\\
      0 & 0 & 1 & 1
  \end{pmatrix}, \quad
  \vec{\pi} = (1,0,0,0), \quad
  \vec{\eta} = \begin{pmatrix} 1\\0\\0\\0 \end{pmatrix}.
\end{equation}

To transform this into a PFA, we need to add a fifth state due to the rows of
zeros, and from the construction we get
\begin{equation}
  P_0' = \begin{pmatrix} 0 & 1 & 0 & 0 & 0\\
    0 & 0 & 1 & 0 & 0\\
    0 & 1/2 & 0 & 1/2 & 0\\
    0 & 0 & 0 & 0 & 1\\
    0 & 0 & 0 & 0 & 1
  \end{pmatrix}, 
  P_1' = \begin{pmatrix} 0 & 0 & 0 & 0 & 1\\
      1 & 0 & 0 & 0 & 0\\
      0 & 0 & 0 & 0 & 1\\
      0 & 0 & 1/2 & 1/2 & 0\\
      0 & 0 & 0 & 0 & 1
  \end{pmatrix}, 
  \vec{\pi}' = (1,0,0,0,0), 
  \vec{\eta}^{\,'} = \begin{pmatrix} 1\\0\\0\\0\\0 \end{pmatrix}.
\end{equation}
However, this is hardly optimal as a witness for $A_P$, since actually
$A_P(0001101)=3$ via
\begin{equation}
  P_0 = \begin{pmatrix} 0 & 1/2 & 1/2\\
    0 & 1/2 & 1/2\\
    0 & 1 & 0
  \end{pmatrix}, \quad
    P_1 = \begin{pmatrix} 0 & 0 & 1\\
      1 & 0 & 0\\
      0 & 1 & 0
    \end{pmatrix}, \quad
    \vec{\pi} = (1,0,0), \quad
    \vec{\eta} = \begin{pmatrix} 1\\0\\0 \end{pmatrix}.
\end{equation}

Indeed, no nonconstant strings are presently known for which $A_P(x)$ is equal
to the maximum possible value $A_N(x)+1$. 
Direct computations have shown that all binary strings $x$ of length 9 or
less have $A_P(x) \leq 3$, whereas many such strings have $A_N(x)=4$ or $5$.
However, unlike with $A_N$, there are no strings $x$ with $A_P(x)=1$, because by
the requirement for matrices to be row-stochastic and for the initial state
distribution to be given by a probability vector, a PFA with one state must
either accept all words with probability $1$ or fail to accept any word. The
only strings with $A_N(x)=1$ are the constant strings $x=a^n$, and we have
$A_P(a^n)=2$ by \thref{classification2-full}.

So far we have not mentioned any examples involving $A_{P,\delta}$.
Experimentally, it would appear that one has to make the value of $\delta$ quite
low in order to get small values of $A_{P,\delta}$, for all but very short
strings. This makes intuitive sense in view of the proof of
\thref{classification2}, and more generally the phenomenon of stability of a
contractive iterated function system: all orbits converge to the attractor, and
correspondingly acceptance probabilities will tend to cluster together for
longer strings, at least for a generic automaton. 
As an example, if $x=0110$, then a simple witness for $A_P(x)=2$ is the PFA $M$
given by
\begin{equation}
  P_0= \begin{pmatrix} 0&1\\ 1/2&1/2 \end{pmatrix},\quad 
  P_1= \begin{pmatrix} 1/2&1/2\\ 0&1 \end{pmatrix},\quad
  \vec{\pi}=(1,0), \quad \vec{\eta}=\begin{pmatrix} 1\\ 0 \end{pmatrix}.
\end{equation}
It follows from the proof of \thref{classification2} that $M$ witnesses
that $A_P(01^m0)=2$ for all $m$, and one can calculate that $\gap_M(0110) =
1/16$, $\gap_M(01^30) = 1/32$, $\gap_M(01^40) = 1/64$, ...

However, it is not necessarily the case that gaps strictly decrease for longer
strings. The PFA
\begin{equation}\label{eq:1n0103}
  P_0 = \begin{pmatrix} 0 & 1 & 0\\
      2/3 & 0 & 1/3\\
    1/3 & 0 & 2/3 \end{pmatrix}, \quad
  P_1 = \begin{pmatrix} 1/3 & 2/3 & 0\\
      2/3 & 1/3 & 0\\
    0 & 0 & 1 \end{pmatrix}, \quad
    \vec{\pi} = (0,0,1), \quad \vec{\eta} = \begin{pmatrix} 1\\0\\0
    \end{pmatrix}
\end{equation}
witnesses that $A_P(1^n010^3) \leq 3$ for all $n\leq 15$, and each of these
strings is given probability $\frac{91}{243}$ with gap $\frac{1}{243}$ (the same
almost certainly holds for all $n$, though we have not yet shown this). This is
unsurprising since $\vec{\pi}$ is the (left) Perron-Frobenius eigenvector of
$P_1$. Of course, such a relation is easily destroyed by perturbing the entries
of $\vec{\pi}$ and $P_1$.

It does not seem very easy to simultaneously make $\delta$ large and
$A_{P,\delta}$ small even for short strings.
One can get a gap of about $0.5621$ for $x$ by using the 3-state PFA
\begin{equation}
  P_0 = \begin{pmatrix} 0 & 0 & 1\\
    0.22151 & 0.77485 & 0.00364\\
    0.9995 & 0 & 0.0005 \end{pmatrix}, 
  P_1 = \begin{pmatrix} 0 & 0.5622 & 0.4378\\
    1 & 0 & 0\\
    0 & 1 & 0 \end{pmatrix}, 
  \vec{\pi} = (1,0,0), \vec{\eta} = \begin{pmatrix} 0\\ 0\\ 1
  \end{pmatrix}.
\end{equation}
But among 2-state PFAs, the highest gap known for $x$ at the time of writing
is approximately $0.1775$, via
\begin{equation}
  P_0 = \begin{pmatrix} 0.16748 & 0.83252\\
    0.98999 & 0.01001 \end{pmatrix}, \quad
  P_1 = \begin{pmatrix} 0.66116 & 0.33884 \\
    0 & 1 \end{pmatrix}, \quad
  \vec{\pi} = (1,0), \quad \vec{\eta} = \begin{pmatrix} 1\\ 0 \end{pmatrix}.
\end{equation}
These automata were found by using the conjugate gradient method to refine the
results of brute-force searching. In the second case, the search was over
roughly 850,000 2-state PFAs and turned up only one giving $x$ a gap greater
than 1/6 (it was about $0.1719$).
Among the same set of PFAs, the largest gap found for $01^30$ was approximately
$0.1178$.

The next result should be compared with the facts that $A_D(xyz) \geq A_D(y)$
and $A_N(xyz)\geq A_N(y)$ for any strings $x,y,z$. (See
\cite[Lemma~12]{KComplexity} and \cite[Theorem~2.4]{H13}. The statement for
$A_N$ can be derived from $A_N(xy)\geq A_N(x)$ and the invariance of $A_N$ under
string reversal.)

\begin{prop}\thlabel{apxy} For all strings $x,y$ and all $\delta \in [0,1)$, we
  have $A_{P, \delta} (xy) \geq A_{P,\delta}(y)$.
\end{prop}
\begin{proof} Given $\delta$, let $M=(S,\Sigma,P,\vec{\pi},\vec{\eta})$ witness
  $A_{P,\delta}(xy)$. 
  Let $M'=(S,\Sigma,P,\vec{\pi}',\vec{\eta})$ be a PFA with the same
  configuration as $M$ except for its initial state distribution, which will be
  $\vec{\pi}' = \vec{\pi}P_M(x)$. Since $P_M(x)$ is a stochastic matrix,
  $\vec{\pi}'$ is still a probability vector. By definition, $P_M(xw) = P_M(x)
  P_M(w)$ for all strings $w$, so we have $\rho_{M'}(w) = \rho_M(xw)$ and
  consequently 
  \begin{equation}
    \gap_{M'}(w) = \min \set{\rho_M(xw)-\rho_M(xz) \st z\in \Sigma^{\abs{w}}
    \setminus \{w\}} \geq \gap_M (xw)
  \end{equation}
  for all $w$.
  In particular, $\gap_{M'}(y) \geq \gap_M(xy) > \delta$, so $M'$ witnesses an
  upper bound for $A_{P,\delta}(y)$.
\end{proof}

One property of $A_N$ which motivated its introduction, as mentioned above, is
that $A_N(x) = A_N( \rev{x})$, where $\rev{x}$ is the reversal of $x$:
\[
  \rev{x} = x_n\cdots x_2x_1 \quad\text{if}\quad x=x_1x_2\cdots x_n.
\]

By \thref{classification2-full}, the class of strings $x$ with $A_P(x)=2$ is not
closed under reversal, so $A_P$ does not share this property.
In Section~\ref{sec:otherdefs} we will take up the question of how one might
recover the property by modifying $A_P$.
Equality of $A_P(x)$ and $A_P(\rev{x})$ is possible in at least some cases:

\begin{prop}\thlabel{reversal} If $A_P(x)$ is witnessed by a PFA $M =
  (S,\Sigma,P, \vec{\pi}, \vec{\eta})$ such that each $P_\sigma\in P$ is doubly
  stochastic, and such that all nonzero entries of $\vec{\pi}$ are equal, then
  $A_P(\rev{x}) \leq A_P(x)$. If $M$ witnesses $A_{P,\delta}(x)$ and one can
  additionally take $\vec{\pi}$ and $\vec{\eta}$ to have the same number of
  nonzero entries, then $A_{P,\delta}(\rev{x}) \leq A_{P,\delta}(x)$.
\end{prop}
\begin{proof} The idea is more or less the content of Exercise A.2.8 of
  Chapter~3 of \cite{Paz71}. Define the PFA $M' = (S,\Sigma, P',\vec{\pi}',
  \vec{\eta}^{\,'})$ by $P'_\sigma = P_\sigma^T$ for each $\sigma\in \Sigma$ and
  $\vec{\pi}'=\vec{\eta}^{\,T}/s$, where $s=\sum \vec{\eta}$. 
  If each entry of $\vec{\pi}$ is either $0$ or $1/n$ (for some $n\geq 1$), then
  let $\vec{\eta}^{\,'} = n\vec{\pi}^T$.

  Intuitively, $M'$ represents the automaton obtained by operating $M$ in reverse.
  We have $P_{M'}(\rev{x}) = P_M(x)^T$, so 
  \begin{align*}
    \rho_{M'}(\rev{x}) = (\vec{\eta}^T/s) P_{M'}(\rev{x}) (n\vec{\pi}^T)
    &= ns\inv\paren{ \vec{\pi} P_M(x)\vec{\eta}}^T\\
    &= ns\inv\paren{ \rho_M(x) }^T = ns\inv\rho_M(x).
  \end{align*}

  The same calculation shows $\rho_{M'}(\rev{y}) = ns\inv \rho_M(y)$ for all
  $y$, so if $\rho_M(x)>\rho_M(y)$, then $\rho_{M'}(\rev{x}) > \rho_{M'}
  (\rev{y})$. Therefore $M'$ witnesses $A_P(\rev{x})\leq A_P(x)$ since $M'$ and
  $M$ have the same number of states. 

  If $\vec{\pi}$ and $\vec{\eta}$ have the same number of nonzero entries, then
  $n=s$, and so $\rho_{M'}(\rev{y}) = \rho_M(y)$ for all $y\in \Sigma^\ast$.
  Hence $\gap_{M'}(\rev{x}) = \gap_M(x)$ and the second statement follows.
\end{proof}

As a corollary of this fact together with \thref{classification2-full} below,
for most binary strings $x$ such that $A_P(x)=2$, the latter cannot be
witnessed by a PFA as in the proposition.
$A_P(\rev{x}) = A_P(x)$ holds whenever both quantities can be witnessed by such
a PFA. An example is given by \eqref{eq:1n0103}, which witnesses $A_P(1^n010^3)
= 3$ and can be turned into a witness for $A_P(0^3101^n)\leq 3$ for all $n
\leq 15$ (at least) using the procedure in the proof of \thref{reversal}. Since
none of the latter family of strings with $n\geq 2$ can have complexity 2 by
\thref{classification2-full}, it follows that $A_P(0^3101^n)=3$ for $2\leq n\leq
15$ (at least).

%% file: classifs.tex
\section{Classification of binary strings with $A_P=2$}\label{sec:class}

This section is devoted to proving the following theorem, restated from the
introduction for the reader's convenience:

\classtwo

This set of strings is significantly larger than the set of binary strings with
NFA complexity 2. As classified in \cite{H13}, the strings with $A_N(w)=2$
consist exactly of
\[
  (ij)^m, \qquad i^mj, \qquad \text{and}\quad ij^m
\]
for all $m$. All that can be generally said about $A_N(i^nj^m)$, for instance,
is that it is no more than $\min\{n,m\}+1$ \cite[Example~4.1]{H13}; indeed it is
unbounded by the proof of \thref{ANAPcor} (for which see
Section~\ref{sec:classafter}).

The proof of this theorem will occupy a substantial portion of
the rest of the paper, and we split it into two halves, the forward and reverse
directions:
\begin{thm}\thlabel{classification2} For a binary string $w$, if $A_P(w)=2$,
  then $w$ is of the form
  \[
    i^nj^m, \qquad i^nj^mi, \qquad i^n(ji)^m, \qquad\text{or}\quad i^nj(ij)^m
    \qquad\text{for some } n,m\geq 0.
  \]
\end{thm}

\begin{thm}\thlabel{classification2-reverse} The values of $A_P(i^nj^m)$,
  $A_P(i^nj^mi)$, $A_P(i^n(ji)^m)$, and $A_P(i^nj(ij)^m)$ are equal to $2$ for
  all $n,m \geq 0$.
\end{thm}

The proof depends on a connection between PFAs and iterated
function systems, and we begin by giving the details of this connection in
Section~\ref{sec:ifs}. Then we prove \thref{classification2} in
Section~\ref{sec:classforward} and \thref{classification2-reverse} in
Section~\ref{sec:classreverse}, ending by collecting some corollaries and
further questions in Section~\ref{sec:classafter}.


\subsection{The iterated function system approach}\label{sec:ifs}

An \bdef{iterated function system} (IFS) on a compact metric space $X$ is a
dynamical system consisting of a finite set of continuous maps $f_1,\dotsc,f_n$
on $X$, viewed as inducing a semigroup action on $X$ under composition. If $X$
is $\R^n$ or a compact subset of it, and the maps $f_i$ are affine maps, then
the IFS is called \bdef{affine}. It is well-known that the attractors of
contractive IFSs are fractals, and the use of affine IFSs for efficient
representation of fractal images has been studied \cite{Barnsley,
CulikDubeRational, Sprott}.

Our interest in IFSs is, for present purposes, limited to the fact that one may
obtain an IFS through the acceptance probability function of a PFA, and in doing
so shed light on the family of strings whose complexity the PFA witnesses.
Connections between IFSs and PFAs are already known: Culik and Dube
\cite{CulikDubeAffine, CulikDubeRational} in effect use PFAs as one method of
generating fractal images, as an alternative to directly employing IFSs. They
also introduce probabilistic affine automata, a generalization of PFAs in which
each input letter corresponds to an affine map to be applied with some
probability. (See \cite{Rystsov} for a more recent study of this idea.)

Koci\'c and Simoncelli in \cite{KS} demonstrated a correspondence
between IFSs given by a set of stochastic matrices and affine IFSs on
lower-dimensional simplices. We present this correspondence in a more elementary
formulation adapted to PFAs, showing that the PFA's acceptance probability
function descends to the IFS in a natural fashion. 
Let $M=(S,\Sigma,P,\vec{\pi},\vec{\eta})$ be a PFA
with $k$ states. If there are $0$ or $k$ accepting states, then $\rho_M$ is
identically $0$ or $1$, respectively, so assume without loss of generality that
there are between $1$ and $k-1$ accepting states. By permuting the states of $M$
(and hence the rows and columns of $\vec{\pi}$, $\vec{\eta}$, and each
$P_\sigma$), we may assume that the $k$th state is not accepting.

Recall that if
$\abs{w}=n$, then $\rho_M(w) = \vec{\pi} P_M(w) \vec{\eta}$, where
$P_M(w)=\prod_{i=1}^n P_{w_{n-i}}$. This just means that $\rho_M(w)$
is a sum of up to $k-1$ elements of the row vector $\vec{\pi}P_M(w)$. We can
think of each multiplication by a $P_\sigma$ as updating the state distribution
$\vec{\pi}$, and of $\vec{\pi}$ itself as representing the state distribution
$\vec{\pi}(\lambda)$ after reading the empty string $\lambda$. Then let
\[
  \vec{\pi}(w) = \paren{p_1(w),p_2(w),\dotsc, p_{k-1}(w), 1-\sum_{i<k}
  p_i(w)} = \vec{\pi}(\lambda)P_M(w)
\]
be the state distribution after reading a string $w$. 
Now, the last component of $\vec{\pi}(w^\smallfrown \sigma)$ only depends on its
first $k-1$ components together with the first $k-1$ columns of $P_\sigma$.
Since the $k$th state of $M$ is not accepting, $\rho_M(w^\smallfrown \sigma)$
thus depends only on the first $k-1$ components of $\vec{\pi}(w)$, and if we
only care about recovering $\rho_M$ then we can drop the $k$th component from
$\vec{\pi}(w)$ without losing any information.

So, let $\vec{a}_i$ be the $i$th row of $P_\sigma$ truncated to its first $k-1$
entries, let $\vec{y}(w) = \vec{\pi}(w) \restr (k-1)$, and let $\vec{1}_{m,n}$
be the $m\times n$ matrix of all $1$s. Also let $U$ be $P_\sigma$ with its last
row and column deleted (so the rows of $U$ are the vectors $\vec{a}_i$ for $i <
k$). Then for any $\sigma\in \Sigma$,
\[
  \vec{y}(w^\smallfrown \sigma) = \vec{y}(w) U + \paren{1 - \sum
    \vec{y}(w)} \vec{a}_k = \vec{a}_k + \vec{y}(w) \paren{U - \vec{1}_{k-1,1}
  \vec{a}_k}.
\]

$\vec{y}(w)$ is an element of the $(k-1)$-dimensional unit simplex $S_{k-1}$, so
we identify $w\mapsto \vec{\pi}(w^\smallfrown \sigma)$ with the map $f_\sigma
\colon S_{k-1}\to S_{k-1}$ that sends $\vec{x}$ to $\vec{a}_k + \vec{x}B$, where
$B =  U - \vec{1}_{k-1,1} \vec{a}_k$. Note that the entries of $B$ may be
negative.
Multiplication by $P_\sigma$ thus corresponds to composition by $f_\sigma$. If
we give the IFS consisting of the functions $f_\sigma$ the starting vector
$\vec{x}_0 = (p_1(\lambda), p_2(\lambda), \dotsc, p_{k-1}(\lambda))$, then we
have 
\begin{equation}\label{eq:rho}
  \rho_M(w) = \sum \set*{ (f_{w_n}\circ f_{w_{n-1}} \circ \cdots \circ f_{w_0}
  (\vec{x}_0))_i : \text{the $i$th state of $M$ is accepting}},
\end{equation}
where $\vec{v}_i$ here denotes the $i$th component of the vector $\vec{v}$ and where
$w=w_0w_1\cdots w_n$. Hence for any $k$-state PFA $M$ there is an affine IFS on
$S_{k-1}$ whose iterations exactly recover the function $\rho_M$ in the above
fashion.

In the other direction, suppose we are given a finite set of affine maps $f_\sigma
\colon \vec{x}\mapsto \vec{a}+\vec{x}B$ on $S_{k-1}$, where $\vec{a}$ and $B$
depend on $\sigma$, along with a starting vector $\vec{x}_0 = (p_1, \dotsc,
p_{k-1})$. We build a PFA $M$ as follows. Let $\tilde{A}$ and $\tilde{B}$ be the
$k\times k$ matrices given by $\tilde{A} = \vec{1}_{k,1} \begin{pmatrix} \vec{a}
& \rvline & 1-\sum \vec{a} \end{pmatrix}$ and
\[
  \tilde{B} = \begin{pmatrix} \vec{1}_{k-1,1}\\ 0 \end{pmatrix} \begin{pmatrix}
  B & \rvline & -B \vec{1}_{k-1,1} \end{pmatrix} = 
  \begin{pmatrix} \begin{matrix} ~ & ~ & ~\\
    ~ & \mbox{\normalfont\Large $B$} & ~\\
    ~ & ~ & ~
  \end{matrix}
  & \rvline & 
  \begin{matrix} - \sum_{i<k} B_{1,i}\\ \vdots\\ -\sum_{i<k} B_{k-1,i}
  \end{matrix}\\
  \hline
    \begin{matrix} ~ & \vec{0} & ~ \end{matrix} & \rvline & \begin{matrix} 0
  \end{matrix}
  \end{pmatrix}.
\]

Then let
\[
  P_\sigma = \tilde{A} + \tilde{B} \quad \text{and} \quad 
  \vec{\pi} = \vec{\pi}(\lambda) = \begin{pmatrix} \vec{x}_0 & \rvline & 1-\sum
  \vec{x}_0 \end{pmatrix} \in \R^k.
\]

Also define
$\vec{\pi}(w)$ for any $w$ as before. Then $P_\sigma$ is
stochastic: first, each row clearly sums to $1$ as the row sums of $\tilde{A}$
and $\tilde{B}$ are all $1$ and $0$, respectively. 
If $\vec{e}_i$ is the $i$th standard basis vector in $\R^{k-1}$, then
$f_\sigma(\vec{e}_i)$ is the sum of $\vec{a}$ and the $i$th row of $B$, i.e.,
the upper left $(k-1)\times (k-1)$ submatrix of $P_\sigma$. From $\vec{a} =
f_\sigma(\vec{0}) \in S_{k-1}$ and $f_\sigma(\vec{e}_i)\in S_{k-1}$ it follows
that each entry of $\tilde{A}+\tilde{B}$ is in $[0,1]$.

One can check that $\vec{\pi}(\lambda) P_\sigma \restr (k-1) = \vec{a} +
\vec{x}_0 B = f_\sigma(\vec{x}_0)$.
Inductively we have that $\vec{\pi}( w^\smallfrown
\sigma) \restr (k-1) = f_\sigma(\vec{x})$ if $\vec{x}= \vec{\pi}(\lambda)
P_M(w)$. Now, the data we have so far does not uniquely specify a PFA $M= (\{
1,\dotsc,k\}, \Sigma, \{P_\sigma\}, \vec{\pi}, \vec{\eta})$, because nothing
about the vector of accepting states $\vec{\eta}$ is implied by the IFS we
started with except that the $k$th state should not be accepting. Thus the same
IFS can be made to correspond to any PFA $M$ having the $\vec{\pi}$ and matrices
$P_\sigma$ given above, and such that the last state is not accepting. The
equation \eqref{eq:rho} holds for any such $M$ and $w$, which completes the
correspondence.

Since we will only apply this correspondence to two-state automata in the
present work, we separately outline this case for clarity. Given a two-state PFA
$M$, write $\vec{\pi}$ as $(p,1-p)$. Assume by permuting the states that
$\vec{\eta} = (1,0)^T$. For each $\sigma\in\Sigma$, write
\begin{equation}\label{eq:psig}
  P_\sigma = \begin{pmatrix} a_\sigma + b_\sigma & 1-a_\sigma-b_\sigma\\
    a_\sigma & 1-a_\sigma\end{pmatrix},
\end{equation}
where $b_\sigma$ is allowed to be negative. Then for each $w\in \Sigma^\ast$, we
have
\[
  \rho(w^\smallfrown \sigma) 
  = \begin{pmatrix} \rho(w) & 1-\rho(w)\end{pmatrix}
   \begin{pmatrix} a_\sigma +
    b_\sigma & 1-a_\sigma-b_\sigma\\
    a_\sigma & 1-a_\sigma\end{pmatrix} \begin{pmatrix} 1\\0 \end{pmatrix}
   = a_\sigma + b_\sigma\rho(w).
\]
We can thus associate with $P_\sigma$ the ``incremental probability function''
\[
  f_\sigma(x) = a_\sigma + b_\sigma x
\]
mapping $[0,1]$ into itself. Viewing $p$ as $\rho(\lambda)$, we obtain the IFS
$(f_\sigma)_{\sigma\in\Sigma}$ with starting value $x_0=p$ such that for any
word $w= w_1w_2\dotsc w_n$,
\begin{equation}\label{eq:2staterho}
  \rho(w) = f_{w_n}\circ f_{w_{n-1}} \circ \cdots \circ f_{w_1} (x_0).
\end{equation}

In the other direction, starting from an IFS given by affine maps $f_\sigma$ on
$[0,1]$ together with $x_0$, setting $\vec{\pi} = (x_0, 1-x_0)$ and defining the
matrices $P_\sigma$ as in \eqref{eq:psig} produces a PFA whose acceptance
probability function satisfies \eqref{eq:2staterho}.

The set of $w$ such that an upper bound for $A_P(w)$ is witnessed by $M$ is
exactly the set of $w$ describing a sequence of compositions maximizing the
value along the orbit of $x_0$ under this IFS. This idea will be exploited
heavily throughout the following section.

%% file: classforward.tex
\subsection{Proof of \thref{classification2}}\label{sec:classforward}

We will establish that any two-state PFA over a binary alphabet must witness the
complexity of only strings in one of the forms given in the theorem, i.e., 
\[
  i^nj^m, \qquad i^nj^mi, \qquad i^n(ji)^m, \qquad\text{or}\quad i^nj(ij)^m,
\]
if it witnesses anything at all. If $i=j$, of course, these strings are constant
and so trivially have complexity 2. Permuting the underlying alphabet does not
change the complexity of a string, as it corresponds merely to permuting the
maps $f_\sigma$ of the IFS (or equivalently the transition matrices $P_\sigma$
of the original PFA). Therefore, any statement in this section about a string
should be understood to apply equally well to its bit-flip (i.e., the result of
permuting 0 and 1), by switching the roles of $f_0$ and $f_1$.

Assume we are given a two-state PFA represented by the IFS
\[
  f_0(x)=a+bx \qquad\text{and}\qquad f_1(x)=c+dx
\]
with starting value $x_0\in [0,1]$, where $f_0$ and $f_1$ map $[0,1]$ into
itself. We use the word ``orbit'' to mean any forward orbit of $x_0$ under the
semigroup action generated by $f_0$ and $f_1$, that is, the orbit of $x_0$ under
some particular sequence of compositions of $f_0$ and $f_1$. We always omit
parentheses when composing functions, so that e.g.\ $f_0^2x = f_0(f_0(x))$. For
brevity, we describe a probability as \bdef{$n$-maximal} (\bdef{$n$-minimal}) if
it is maximal (minimal) among the probabilities of strings of length $n$. We
also refer to an $n$-maximal ($n$-minimal) probability as simply an $n$-maximum
($n$-minimum). If $n$ is clear from context, we may call such a probability
maximal (minimal) or a maximum (minimum). We say the IFS witnesses a string $w$
if it witnesses that $A_P(w)=2$, i.e., $\rho(w)$ is maximal. The following
elementary observations will be useful throughout:

\begin{itemize}
  \item For $i\in \{0,1\}$, if $f_i$ is not the line $y=x$, then $f_i$ has a
    unique fixed point in $[0,1]$, towards which it contracts with rate equal
    to the absolute value of its slope. The case $y=x$ will be dispensed with in
    \thref{classobs}, and we can assume elsewhere that neither $f_0$ nor $f_1$
    is the identity map. We will use $r_0$ and $r_1$ to denote the fixed points
    of $f_0$ and $f_1$, respectively. By abuse of notation, $r_0$ and $r_1$
    refer either to the $x$-coordinates of these points or to the actual points
    in $[0,1]^2$. It will be clear which is meant from the context. We have
    $r_0=a/(1-b)$ and $r_1=c/(1-d)$.
  \item If $f_i$ has positive slope, then it maps $[0,r_i)$ into itself and
    $(r_i,1]$ into itself. If it has negative slope, it maps $[0,r_i)$ into
    $(r_i,1]$ and vice versa. (If its slope is $0$, of course, it sends every
    point to $r_i$.)
  \item If a probability $x$ is $n$-maximal, then $x$ is either the image of an
    $(n-1)$-maximum under a map of positive slope, or the image of an
    $(n-1)$-minimum under a map of negative slope. Hence we need only consider
    the maximum and minimum probabilities of each length in order to determine
    the maximal-probability strings.
  \item Suppose $f_0$ and $f_1$ intersect at the single point $(i_x,i_y) \in
    [0,1]^2$, and that the maximum or minimum probability of some length turns
    out to equal $i_x$. (We always assume the maps do not coincide, since no
    strings can be witnessed if they do.) Then no further probabilities in the
    same orbit can be unique (since $f_0i_x = f_1i_x$). In this case, no further
    strings are witnessed if their probabilities are in the same orbit as $i_x$.
    We assume for simplicity that this does not happen in the arguments that
    follow. This does not lose any generality, because if $i_x$ happens to be
    attained as the maximal or minimal probability in some orbit, then nothing
    changes about the behavior of the IFS except for the lack of uniqueness of
    the subsequent maxima and minima.
\end{itemize}

For clarity, we separate the argument into several progressively more
complicated cases based on the signs of $b$ and $d$. First we quickly dispense
with some easy ones: if both $f_0$ and $f_1$ are constant, the PFA witnesses
either $0^n$ or $1^n$ for all $n$, depending on which line is higher. If one
map is the identity, or more generally if $f_0$ and $f_1$ commute, no strings
can be witnessed beyond constant strings, because the only determining factor of
$\rho(w)$ is the number of $0$s and $1$s in $w$.

The following lemma collects various facts which will be useful
for the rest of the proof.
\begin{lem}\thlabel{classobs} Let $f_0=a+bx$ and $f_1=c+dx$ be maps from the
  unit interval into itself. Assume by convention that $a<c$ and that the maps
  intersect at the unique point $(i_x,i_y)\in [0,1]^2$. Taken together, these
  imply in particular that $b>d$, and we will always assume $a<c$ and $b>d$.
  \begin{enumerate}[label=(\alph*),leftmargin=*]
  \item If neither map is the identity, then either both maps fix $i_x$ or
    neither does. We always assume neither map is the identity from this point
    on, as well as that the maps do not coincide.
  \item Both maps fix $i_x$, i.e., $i_y=i_x$, iff $r_0=r_1=i_x$ iff
    $f_0f_1=f_1f_0$. 
  \item \label{lemmapartc} Both maps decrease $i_x$, i.e., $i_y<i_x$, iff
    $r_0<r_1<i_x$ iff $f_0f_1<f_1f_0$ iff $f_0f_1^2<f_1^2f_0$. 
  \item Both maps increase $i_x$, i.e., $i_y>i_x$, iff $r_0>r_1>i_x$ iff
    $f_1f_0<f_0f_1$ iff $f_1^2f_0<f_0f_1^2$.
  \item \label{lemmaparte} If $f_0$ and $f_1$ both have negative slopes, and if
    neither fixes $i_x$, then $\abs{r_0-r_1}<\abs{r_1-i_x}$. 
  \item \label{lemmapartf} Suppose $f_0$ and $f_1$ both have negative slopes. If
    both maps decrease $i_x$, then if $x\in [r_0,i_x)$, every orbit of $x$
    remains inside $[0,i_x)$. If both maps increase $i_x$, then if $x\in
    (i_x,r_0]$, every orbit of $x$ remains inside $(i_x,1]$.
\end{enumerate}
\end{lem}
\begin{proof} \begin{enumerate}[label=(\alph*),leftmargin=*]
  \item $r_0$ and $r_1$ are the intersections of $f_0$ and $f_1$, respectively,
    with $y=x$. If (say) $r_0=i_x$, then $(i_x,i_x)$ also lies on $y=x$ and is
    in the range of $f_1$, therefore $r_1=i_x$ too.

  \item The first equivalence is obvious by definition. Notice $f_0f_1x =
    a+bc+bdx$ and $f_1f_0x= c+ad+bdx$. Then
    \[
      f_0f_1=f_1f_0 \iff a+bc=c+ad \iff a/(1-b) = c/(1-d), 
    \]
    i.e., $f_0f_1 = f_1f_0$ iff $r_0=r_1$, and this happens iff they both
    equal $i_x$ since $r_0$ and $r_1$ both lie on the line $y=x$.

  \item Both $r_0$ and $r_1$ are less than $i_x$ in this case, because the maps
    contract towards their fixed points, so if $f_ix < x$ then $x>r_i$. We have
    $i_x=(c-a)/(b-d)$ and $i_y= (bc-ad)/(b-d)$. Remembering that our assumptions
    imply $b>d$ no matter the sign of each, and observing that neither $b$ nor
    $d$ equals $1$, we have $i_x>i_y$ iff
    \begin{gather*}
      \frac{c-a}{b-d} > \frac{bc-ad}{b-d} \iff c-a>bc-ad \iff c(1-b)>a(1-d)\\
      \iff \frac{c}{1-d}>\frac{a}{1-b} \iff r_1>r_0.
    \end{gather*}
    Since $r_1>r_0 \iff c+ad > a+bc \iff f_1f_0>f_0f_1$, this completes the
    second equivalence. For the third, note $f_0f_1^2x = a+b(c+cd+d^2x)$ and
    $f_1^2f_0x = c+cd+d^2(a+bx)$. Then 
    \begin{gather*}
      f_0f_1^2x < f_1^2f_0x \iff a+bc + bcd +bd^2x < c+cd+ad^2 +bd^2x \\
      \iff
      d(bc-ad) < (c-a) - c(b-d) \iff c + d \frac{bc-ad}{b-d} < \frac{c-a}{b-d}\\
      \iff c + di_y < i_x \iff f_1i_y < i_x \iff f_1^2i_x < i_x.
    \end{gather*}
    The last inequality holds iff $r_1<i_x$ (as $f_1^2$ contracts $i_x$ towards
    its fixed point $r_1$), which happens iff $i_x>i_y$ by the first
    equivalence.

  \item This follows by swapping ``$>$'' with ``$<$'' everywhere in the argument
    for part \ref{lemmapartc}.

  \item By writing $c=r_1(1-d)$, one can rearrange the formula $i_y=c+di_x$ to
    obtain $d=(i_y-r_1)/(i_x-r_1)$. Since $\abs{d}\leq 1$, this implies
    $\abs{i_y-r_1}\leq \abs{i_x-r_1}$. We will finish the proof by showing that
    when $b<0$, then in fact $i_x > i_y$ iff $i_y<r_0$ and $i_x<i_y$ iff
    $i_y>r_0$. That is, depending on whether both maps decrease or increase
    $i_x$, we have either $i_y<r_0 < r_1 < i_x$ or $i_y>r_0>r_1>i_x$. Then
    \[
      \abs{i_y-r_1}=\abs{i_y-r_0} + \abs{r_0- r_1} > \abs{r_0-r_1},
    \]
    and one gets $\abs{r_0-r_1} < \abs{i_y-r_1} \leq \abs{i_x-r_1}$.

    So, $i_y<r_0$ if and only if
    \begin{gather*}
      \frac{bc-ad}{b-d} < \frac{a}{1-b} \iff (bc-ad)(1-b) < a(b-d) \iff bc-b^2c
      + abd < ab\\
      \iff bc(1-b) < ab(1-d) \iff c(1-b) > a(1-d) \iff \frac{c}{1-d}>
      \frac{a}{1-b},
    \end{gather*}
    i.e., if and only if $r_1>r_0$, which is equivalent to $i_x>i_y$. (The
    change from $<$ to $>$ in the second line is because $b<0$.) It is clear
    that one can switch ``$<$'' and ``$>$'' everywhere in this argument to
    obtain that $i_y>r_0$ iff $i_x<i_y$, and the proof is complete.

  \item For the first claim, by assumption $i_y<i_x$ and so $r_0<r_1<i_x$. 
    Since $f_0i_x=f_1i_x$, we have $f_0f_1i_x = f_0f_0i_x$, which is less than
    $i_x$. This implies that $r_{01}$, the fixed point of $f_0f_1$, is also less
    than $i_x$: if $f_0f_1$ decreases the value of a point, then that point must
    be above $r_{01}$. As $f_0f_1$ contracts to $r_{01}$, we have that $f_0f_1x
    < i_x$ whenever $x<i_x$. In other words, if $\rho(w)<i_x$, then
    $\rho(w^\smallfrown 10)<i_x$ too. The analogous statement holds for
    $f_1f_0$, i.e., $\rho(w) < i_x$ implies $\rho(w^\smallfrown 01)<i_x$.
    Finally, since $\abs{r_0 - r_1} < \abs{r_1-i_x}$ by part \ref{lemmaparte},
    $f_1$ always sends points in $[r_0,i_x)$ to points below $i_x$ (and of
    course the same statement is clearly true for $f_0$). This is clear if $x\in
    [r_1,i_x)$. If $x\in [r_0,r_1)$, then $\abs{f_1x-r_1} < \abs{x-r_1} <
    \abs{r_0-r_1} < \abs{i_x-r_1}$, so $f_1x$ is closer to $r_1$ than $i_x$ is,
    and must be less than $i_x$. Overall, then, we have that once an orbit
    enters $[r_0,i_x)$, it stays below $i_x$. The second claim can be proven by
    switching ``$<$'' and ``$>$'' everywhere in the above argument.
\end{enumerate}
\end{proof}

Now begins the main body of the proof of \thref{classification2}. It is split
into four cases: the maps do not intersect, they intersect and have positive
slope, they intersect and have negative slope, and they intersect with one
having positive and the other having negative slope. The last three cases are
each split into two further subcases, based on whether the maps increase or
decrease $i_x$.

\noindent \textbf{Case 1:} $f_0$ and $f_1$ do not intersect in $[0,1]$. Suppose
without loss of generality that $f_1x > f_0x$ for all $x\in [0,1]$, so that
$a<c$. If both maps have positive slope, it follows that $\rho(1^n)$ is maximal
for all $n$. If both have negative slope, then appending $0$ to a maximal
probability always leads to a minimal probability, and appending $1$ to a
minimal probability always leads to a maximal probability. Therefore $(01)^n$
and $1(01)^n$ are witnessed for all $n$, as $f_0x_0 < f_1x_0$. If $f_1$ has
positive and $f_0$ has negative slope, $1^n$ is witnessed for all $n$; note the
ranges of $f_0$ and $f_1$ cannot overlap here. And if $f_1$ has negative and
$f_0$ positive slope, then $0^n1$ is witnessed for all $n$: a minimum can only
be reached by adding all $0$s, again because the ranges are disjoint, and since
$f_0f_1x < f_1f_0x$ for all $x$ (\thref{classobs}\ref{lemmapartc}), appending
$10$ to a string always gives a lower probability than appending $01$ does.

\noindent\textit{Strings witnessed in this case:} $1^n$,
$(01)^n$, $1(01)^n$, $0^n1$.

\noindent \textbf{From now on, assume} that $a<c$, that $f_0$ and $f_1$
intersect at the point $(i_x, i_y)\in [0,1]^2$, and that the maps do not
commute. By \thref{classobs}, this implies that $r_0$, $r_1$, and $i_x$ are
distinct. Assuming $a<c$ is no loss of generality, because we are working only
up to permuting $f_0$ and $f_1$, and $a<c$ is required for the maps to intersect
within the unit interval. As noted in the lemma, this also implies $b>d$ in
every case.

\begin{figure}
\centering
\begin{subfigure}[b]{0.4\textwidth}\begin{adjustbox}{width=\linewidth}
  \input{pictures-pos2.tex}
\end{adjustbox}
\caption{Both maps decrease $i_x$}
\end{subfigure}
\begin{subfigure}[b]{0.4\textwidth}\begin{adjustbox}{width=\linewidth}
  \input{pictures-pos3.tex}
\end{adjustbox}
\caption{Both maps increase $i_x$}
\end{subfigure}
\caption[Possible subcases of Case~2 in the proof of \thref{classification2},
where both maps in the IFS have positive slope.]{Subcases for $f_0$, $f_1$ with positive slope (Case 2)}
\label{fig:positive}
\end{figure}
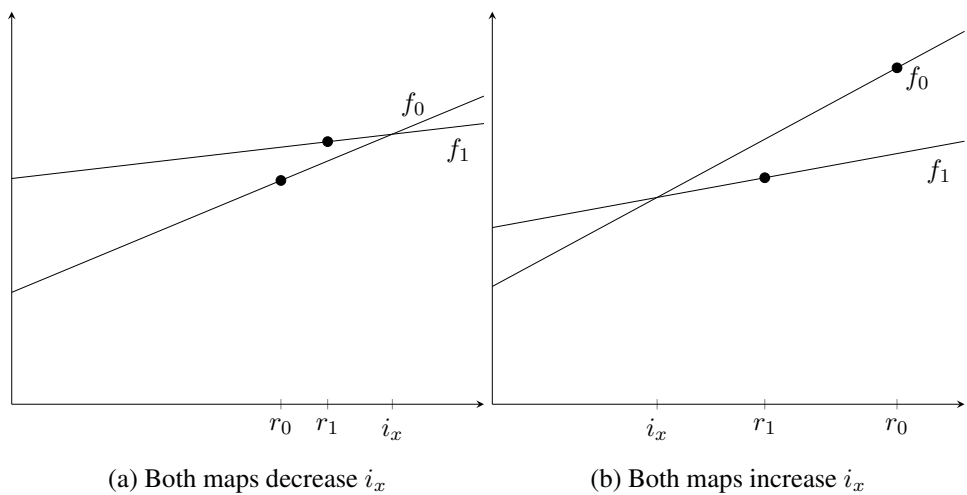

\noindent \textbf{Case 2:} both $f_0$ and $f_1$ have nonnegative slopes.
Specifically, assume $a,b,c\geq0$ and $d>0$ (the case $b=d=0$ is trivial).
There are then two possible subcases of this case, which are illustrated in
Figure~\ref{fig:positive}:
\begin{enumerate}[label=(\alph*),leftmargin=*]
  \item Both $f_0$ and $f_1$ decrease $i_x$. Then we must have $r_0<r_1<i_x$,
    and $0^{n_0}1^m$ is witnessed for all $m$, where $n_0 \geq0$ is least such
    that $f_0^{n_0}x_0 < i_x$ (taking $f_0^0$ to be the identity map). This is
    because $f_0x>f_1x$ for $x>i_x$, but iterating it will eventually cause the
    value to drop below $i_x$, and from that point on, $f_1>f_0$. If $x_0<i_x$
    then we have $f_1>f_0$ from the start.
  \item Both $f_0$ and $f_1$ increase $i_x$. Then $i_x<r_1<r_0$, and
    $1^{n_0}0^m$ is witnessed for all $m$, where $n_0\geq0$ is least such that
    $f_1^{n_0}x_0 > i_x$. The reasoning is exactly analogous to that in (a).
\end{enumerate}
\textit{Strings witnessed in the above case:} $1^n$, $0^n1^m$, $1^n0^m$.

\noindent \textbf{Case 3:} both $f_0$ and $f_1$ have negative slopes. The special
case $b=0$ and $d<0$ is discussed under Case 4 below, so assume $b$ and $d$ are
both strictly negative here. Recall that having negative slopes means each $f_i$
``flips $x$ over'' $r_i$: if $x<r_i$, then $f_ix>r_i$, and vice versa. If we
start an orbit with a maximal probability of its length, then the orbit can only
lead to maximal probabilities for odd-length strings (and this is the only way
to witness odd-length strings). This is accomplished by extending a string on
even lengths in order to achieve a \emph{minimal} probability. For the same
reason, if we start an orbit with a minimal probability, then only even-length
strings may have maximal probabilities in that orbit. The two essentially
different cases for the configuration of $f_0$ and $f_1$ are shown in
Figure~\ref{fig:negative}.

\begin{itemize}[leftmargin=*]
  \item[(a)] Both $f_0$ and $f_1$ decrease $i_x$. Then $r_0<r_1<i_x$. Recall
    that by \thref{classobs}\ref{lemmapartf}, once an orbit enters $[r_0,i_x)$,
    it stays below $i_x$ forever.

    Suppose $x_0>i_x$ and we start an orbit with the maximal-probability string
    $0$. Then $\rho(0)<i_x$, and $\rho(0^2)$ is minimal. If $\rho(0^2)>i_x$,
    then $\rho(0^3)$ is maximal, since $f_0>f_1$ above $i_x$. Every
    $(2\ell+1)$-maximal probability is an image of a $2\ell$-minimal
    probability, so as long as $\rho(0^{2\ell})>i_x$, we have that
    $\rho(0^{2\ell+1})$ is maximal and $\rho(0^{2\ell+2})$ is minimal. Let $n_0$
    be least such that $\rho(0^{2n_0}) < i_x$. Once this happens, since
    $\rho(0^{2n_0})$ is minimal and $f_0<f_1$ below $i_x$, $\rho(0^{2n_0}1)$ is
    maximal.

    $\rho(0^{2n_0})$ is between $r_0$ and $i_x$, so we know that its future orbit
    will always stay below $i_x$ by \thref{classobs}\ref{lemmapartf}. This means
    that a maximum is always reached by appending $1$ to a minimum, and a
    minimum is always reached by appending $0$ to a maximum. Hence, among
    odd-length strings with length greater than $2n_0+1$, we witness
    $0^{2n_0}1(01)^m$ for all $m\geq0$. 

    Next, say $x_0>i_x$ and we start our orbit with the minimal-probability
    string $1$. Then $\rho(11)$ is maximal. If $\rho(11)>i_x$, then $\rho(1^3)$
    is minimal and $\rho(1^4)$ is maximal. So we initially witness $1^{2\ell}$
    for $\ell\leq n_0$, where $n_0$ is least such that $\rho(1^{2n_0}) < i_x$.
    Among longer strings, we then witness $1^{2n_0}(01)^m$ for all $m$. To see
    this, one argues in a similar way as when $x_0<i_x$, since $f_1f_0x<i_x$
    when $x\in [r_0,i_x)$. The only difference is that once $\rho(1^{2n_0})
    <i_x$, the $(2n_0+1)$-minimal probability is attained by $\rho(1^{2n_0}0)$,
    as $f_0x<f_1x$ for $x<i_x$. Then appending a $1$ gives the
    $(2n_0+2)$-maximum $\rho(1^{2n_0}01)$, and we continue appending $01$ to
    keep the min-max pattern going and get $(2n_0+2k)$-maximal probabilities for
    all $k$. This concludes the subcase $x_0>i_x$.

    Finally, suppose $x_0<i_x$. This is analogous to the case $x_0>i_x$, but
    with even-odd parity swapped everywhere. In fact, we only need consider the
    case $x_0<r_0$, because when $x_0\in [r_0,i_x)$, we know that all orbits
    stay below $i_x$, so for such $x_0$ we witness $(10)^m$ and $0(10)^m$ for
    all $m\geq0$.

\begin{figure}
\centering
\begin{subfigure}[b]{0.4\textwidth}\begin{adjustbox}{width=\linewidth}
  \input{pictures-neg2.tex}
\end{adjustbox}
\caption{Both maps decrease $i_x$}
\end{subfigure}
\begin{subfigure}[b]{0.4\textwidth}\begin{adjustbox}{width=\linewidth}
  \input{pictures-neg3.tex}
\end{adjustbox}
\caption{Both maps increase $i_x$}
\end{subfigure}
\caption[Possible subcases of Case~3 in the proof of \thref{classification2},
where both maps have negative slope.]{Subcases for $f_0$, $f_1$ with negative slope (Case 3)}
\label{fig:negative}
\end{figure}
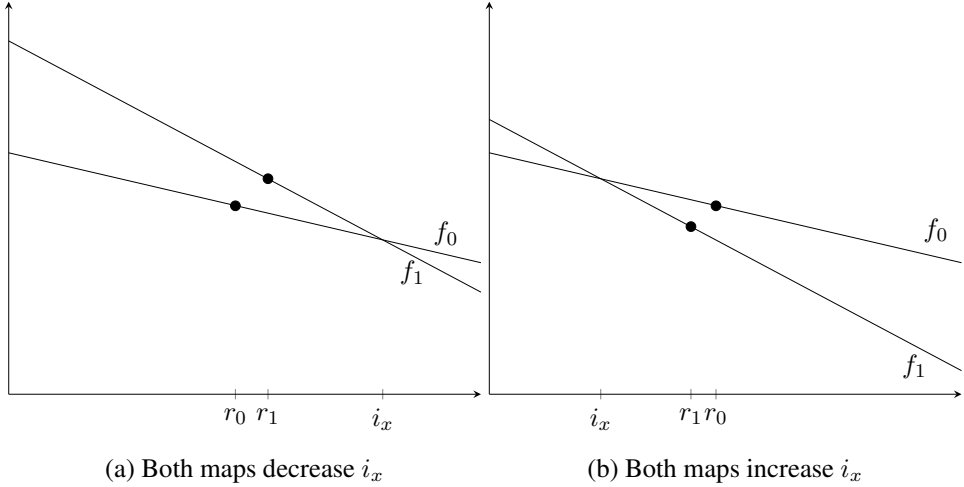

Now, if $x_0<r_0$ and we start with the maximal-probability string $1$, we at
first witness $1^{2\ell+1}$ among odd-length strings, as long as
$\rho(1^{2\ell+1}) > i_x$. If $n_0$ is least such that $\rho(1^{2n_0+1}) < i_x$,
then we witness $1^{2n_0+1}$ and subsequently $1^{2n_0+1}(01)^m$ for all
$m\geq1$. This is because once $\rho(1^{2n_0+1})<i_x$, then the
$(2n_0+2)$-minimum is $\rho(1^{2n_0+1}0)$, followed by the $(2n_0+3)$-maximum
$\rho(1^{2n_0+1}01)$, and continuing to append $01$ keeps the min-max pattern
going. Starting instead with the minimal-probability $0$, we witness
$0^{2\ell+2}$ among even-length strings as long as $\rho(0^{2\ell+1})>i_x$.  If
$n_0$ is least such that $\rho(0^{2n_0+1}) < i_x$, then $\rho(0^{2n_0+1})$ is
minimal but $\rho(0^{2n_0+2})$ is not maximal. Therefore $\rho(0^{2n_0 + 1}1)$
must be maximal, and we witness $0^{2n_0+1}1(01)^m$ for all $m\geq0$. The
pattern of appending $01$ can be repeated forever to obtain maximal
probabilities because $\rho(0^{2n_0 + 1}1) \in [r_0,i_x)$ and applying $f_1f_0$
will always stay below $i_x$, where $f_0$ is minimal and $f_1$ is maximal.
\end{itemize}

\noindent \textit{Strings witnessed in the above case:} $0^{2m}$, $1^{2m+1}$,
$0^{2n}1(01)^m$, $1^{2n}(01)^m$, $0^{2n+1}1(01)^m$.

\begin{itemize}[leftmargin=*]
  \item[(b)] Both $f_0$ and $f_1$ increase $i_x$. Then $i_x<r_1<r_0$. By
    \thref{classobs}\ref{lemmapartf}, once an orbit enters $(i_x,r_0]$, it stays
    above $i_x$.

    Let $x_0<i_x$. By starting with the maximal $\rho(1)$, we have that
    $\rho(1^{2\ell+1})$ is maximal as long as $\rho(1^{2\ell}) < i_x$. (Note
    that $\rho(1^{2\ell+1})$ is always greater than $r_1$ and hence also $i_x$.)
    If $n_0$ is least such that $\rho(1^{2n_0}) > i_x$, then $\rho(1^{2n_0})$ is
    $2n_0$-minimal but $\rho(1^{2n_0+1})$ is not $(2n_0+1)$-maximal. Therefore
    $\rho(1^{2n_0}0)$ is $(2n_0+1)$-maximal, and since $\rho(1^{2n_0}) \in
    (i_x,r_0]$, we have that $\rho(1^{2n_0}0(10)^m)$ remains above $i_x$ for all
    $m\geq0$ and is therefore maximal. By starting instead with the minimal
    $\rho(0)$, we have that $\rho(0^{2\ell+2})$ is maximal as long as
    $\rho(0^{2\ell}) < i_x$. If $n_0$ is least such that $\rho(0^{2n_0}) > i_x$,
    then $\rho(0^{2n_0})$ is $2n_0$-maximal but $\rho(0^{2n_0+1})$ is not $(2n_0
    + 1)$-minimal, since $f_0>f_1$ for $x> i_x$. Therefore $\rho(0^{2n_0}1)$ is
    minimal, and because $\rho(0^{2n_0}) \in (i_x,r_0]$, its future orbits stay
    above $i_x$ and we have $\rho(0^{2n_0}(10)^m)$ maximal for all $m\geq0$.

    If $x_0 \in (i_x,r_0]$, then we witness $(10)^m$ and $0(10)^m$ for all $m
    \geq 0$, as in case (a) when $x_0 \in [r_0,i_x)$. If $x_0>r_0$ and we start
    with the maximal $\rho(0)$, then $\rho(0^{2\ell+3})$ is maximal and
    $\rho(0^{2\ell+2})$ is minimal as long as $\rho(0^{2\ell+1})<i_x$. If $n_0$
    is least such that $\rho(0^{2n_0+1}) > i_x$, then $\rho(0^{2n_0+1})$ is
    $(2n_0+1)$-maximal but $\rho(0^{2n_0 + 2})$ is not $(2n_0+2)$-minimal.
    Instead, $\rho(0^{2n_0+1}1)$ is $(2n_0+2)$-minimal, and since
    $\rho(0^{2n_0+1}) \in (i_x,r_0]$, all future orbits stay above $i_x$, where
    $f_0>f_1$. Therefore $\rho(0^{2n_0+1}(10)^m)$ is maximal for all $m\geq0$.
    If instead we start with the minimal $\rho(1)$, then $\rho(1^{2\ell + 2})$
    is maximal as long as $\rho(1^{2\ell+1}) < i_x$. If $n_0$ is least such that
    $\rho(1^{2n_0+1})> i_x$, then $\rho(1^{2n_0+1})$ is $(2n_0+1)$-minimal but
    $\rho(1^{2n_0+2}) < \rho(1^{2n_0+1}0)$, which is now $(2n_0+2)$-maximal.
    Since $\rho(1^{2n_0+1}) \in (i_x, r_0]$, all of its future orbits stay above
    $i_x$, and thus $\rho(1^{2n_0+1}0(10)^m)$ is maximal for all $m\geq0$.

\end{itemize}

\noindent\textit{Strings witnessed in the above case:} $1^{2m+1}$, $1^{2n}0(10)^m$,
    $0^{2n+1}(10)^m$, $1^{2n+1}0(10)^m$.

\noindent \textbf{Case 4:} $f_0$ has positive slope and $f_1$ has negative
slope. The basic possibilities are illustrated in Figure~\ref{fig:mixed}.

\begin{itemize}[leftmargin=*]
  \item[(a)] Both $f_0$ and $f_1$ decrease $i_x$. \thref{classobs} implies that
    this is equivalent to $f_1f_0x > f_0f_1x$ for all $x$, so that appending
    $01$ always gives a higher probability than appending $10$ would to the same
    string. Assume for now that $b>0$; we will treat the special case $b=0$
    below. The general pattern when $x_0>i_x$ will follow from the next three
    claims:
  \begin{claim}\thlabel{claim1} There is an $n_0$ such that $\rho(1^{2n_0+1}) \geq \rho(1^{
    2n_0-1}0^2)$.
  \end{claim}
  \begin{proof} The map $f_1$ contracts to $r_1$, which is greater than $r_0$. Therefore
    $\rho(1^{2n+1})\geq r_0$ for some $n$. If that is the case, then either
    $\rho(1^{2n-1})\leq r_0$ and so is $\rho(1^{2n-1}0^2)$, or if
    $\rho(1^{2n-1})\geq r_0$, then appending $0^2$ to $1^{2n-1}$ decreases the
    probability towards $r_0$ while appending $1^2$ increases it towards $r_1$.
  \end{proof}
  From now on, take $n_0$ to be the least value as in the previous claim. If
  $x_0>i_x$, then $0$ is 1-maximal, and it follows that $n_0\geq 1$. If
  $\rho(1^{2n_0+1}) = \rho(1^{2n_0-1}0^2)$, then both are minimal, so
  no further strings will be witnessed as there are no longer unique minima or
  maxima of any greater length. Hence without loss of generality assume the
  inequality is strict. The second and third claims will apply to the case
  $n_0>1$; the case $n_0=1$ is handled separately afterwards.

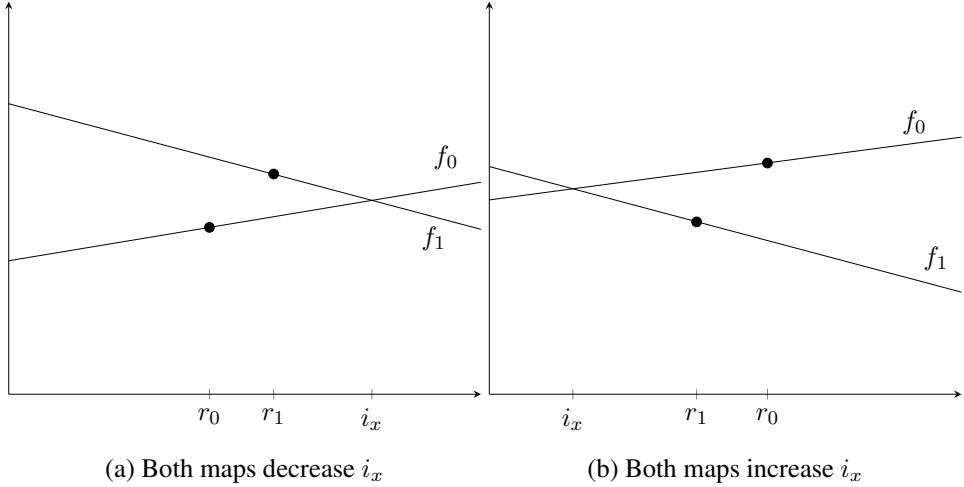
\begin{figure}
\centering
\begin{subfigure}[b]{0.4\textwidth}\begin{adjustbox}{width=\linewidth}
  \input{pictures-mix2.tex}
\end{adjustbox}
\caption{Both maps decrease $i_x$}
\end{subfigure}
\begin{subfigure}[b]{0.4\textwidth}\begin{adjustbox}{width=\linewidth}
  \input{pictures-mix3.tex}
\end{adjustbox}
\caption{Both maps increase $i_x$}
\end{subfigure}
\caption[Possible subcases of Case~4 in the proof of \thref{classification2},
where one map has positive slope and the other has negative slope.]{Subcases for $f_0$ with positive and $f_1$ with negative slope (Case 4)}
\label{fig:mixed}
\end{figure}

  \begin{claim}\thlabel{claim2} If $n_0>1$, then for all $1\leq \ell\leq n_0$, we have:
    \begin{enumerate}[label=(\greek*)]
      \item $\rho(1^{2\ell})$ is $2\ell$-maximal,
      \item $\rho(1^{2\ell-1}0)$ is $2\ell$-minimal,
      \item $\rho(1^{2\ell-1}01)$ is $(2\ell+1)$-maximal, and
      \item if $\ell < n_0$, then $\rho(1^{2\ell+1})$ is $(2\ell+1)$-minimal.
    \end{enumerate}
  \end{claim}
  \begin{proof} By induction on $\ell$. For $\ell=1$, because $\rho(0) >
    \rho(1)$, the only possible 2-maxima are $\rho(00)$ and $\rho(11)$. If we
    have $\rho(11)< \rho(00)$, then because $f_1x < f_1y$ iff $x>y$ for any
    $x,y$, also $f_1^3x_0 > f_1f_0^2x_0$. But from $f_0f_1<f_1f_0$ it follows
    that $f_1f_0^2x_0 > f_0f_1f_0x_0 > f_0^2f_1x_0 = \rho(10^2)$, as $f_0x <
    f_0y$ iff $x<y$ for any $x,y$. (The latter is due to $f_0$ having positive
    slope.) Therefore $\rho(1^3)> \rho(10^2)$, or in other words $n_0=1$. Since
    we are assuming $n_0>1$, this is a contradiction, hence $\rho(00)<\rho(11)$
    and the latter is 2-maximal, which establishes the base case of $(\alpha)$.

   Next, because $f_1f_0>f_0f_1$, we have $\rho(01)> \rho(10)$. The latter is
   less than $\rho(00)$: $\rho(1)<\rho(0)$, so if $\rho(1)<r_0$ then
   $\rho(10)<r_0<\rho(00)$. If $\rho(1)\geq r_0$, then appending a 0 moves
   $\rho(10)$ closer to $r_0$ than $\rho(00)$ is, i.e., makes it smaller than
   $\rho(00)$. This implies $(\beta)$ holds for $\ell=1$. Then $(\gamma)$
   follows if $(\alpha)$ and $(\beta)$ hold for any $\ell$: the only possible
   candidates for a $(2\ell + 1)$-maximum are $\rho(1^{2\ell}0)$ and
   $\rho(1^{2\ell-1}01)$, i.e., the image of the $2\ell$-maximum under $f_0$ and
   the image of the $2\ell$-minimum under $f_1$. But $\rho(1^{2\ell}0) <
   \rho(1^{2\ell-1}01)$ since $f_0f_1 < f_1f_0$. For $(\delta)$, suppose
   $(\alpha)$ and $(\beta)$ are true for $\ell$ and $\ell<n_0$. Then only
   $\rho(1^{2\ell-1}0^2)$ or $\rho(1^{2\ell + 1})$ could possibly be minima,
   since they are the images of the $2\ell$-minimum under $f_0$ and the
   $2\ell$-maximum under $f_1$, respectively. And we have $\rho(1^{2\ell+1}) <
   \rho(1^{2\ell - 1}0^2)$ because $\ell<n_0$.

   Now suppose all four items hold for some given $\ell<n_0$. Then if $(\alpha)$
   and $(\beta)$ hold for $\ell+1$, so does $(\gamma)$ by the above argument,
   and if $\ell + 1<n_0$ then additionally $(\delta)$ holds for $\ell+1$. Hence,
   for the inductive step, it only remains to establish that $(\alpha)$ and
   $(\beta)$ hold for $\ell+1$. For $(\alpha)$, because a $(2\ell+2)$-maximum is
   either the image under $f_1$ of a $(2\ell+1)$-minimum or the image under
   $f_0$ of a $(2\ell+1)$-maximum, the only possible $(2\ell+2)$-maxima are
   $\rho(1^{2\ell + 1}1)$ and $\rho(1^{2\ell-1}010)$. But we have
   $\rho(1^{2\ell-1}010) < \rho(1^{2 \ell-1}001)$ because $f_0f_1<f_1f_0$, so
   $\rho(1^{2\ell-1}010)$ is not maximal. Finally, for $(\beta)$, $\rho(
   1^{2\ell+1}0)$ is $(2\ell+2)$-minimal because the only other possible
   candidate for a minimum is $\rho(1^{2\ell-1}011)$, and $\rho(1^{2\ell+1}0)$ is
   less than $\rho(1^{2\ell-1}011)$. The latter follows by \thref{classobs}, as
   $i_x>i_y$ if and only if $f_0f_1^2 < f_1^2f_0$, so that appending $110$
   always results in a lower probability than appending $011$ to the same
   string.
  \end{proof}

  \begin{claim}\thlabel{claim3} If $n_0>1$, then $\rho(1^{2n_0-1}0^m)$ is $(2n_0-1+m)$-minimal,
    and hence $\rho(1^{2n_0-1}0^m1)$ is $(2n_0+m)$-maximal, for all $m \geq 0$.
  \end{claim}
  \begin{proof} The cases $m=0$ and $m=1$ are covered by taking $\ell=n_0-1$ and
    $\ell=n_0$ in the previous claim. If $\rho(1^{2n_0-1}0^m)$ is
    $(2n_0-1+m)$-minimal, and $\rho(1^{2n_0 - 1}0^{m-1}1)$ is
    $(2n_0-1+m)$-maximal, then only $\rho(1^{2n_0 - 1}0^m0)$ or
    $\rho(1^{2n_0-1}0^{m-1}11)$ could be $(2n_0+ m)$-minimal. But $f_0f_1^2 <
    f_1^2f_0$ implies $\rho(1^{2n_0-1} 0^{m-2}110) < \rho(1^{2n_0-1}0^{m-1}11)$,
    so the latter is not minimal. Finally, this implies
    $\rho(1^{2n_0-1}0^{m+1}1)$ is $(2n_0+m+1)$-maximal: the only other
    possibility is $\rho(1^{2n_0-1}0^{m-1}10)$, and this is less than $\rho(1^{
    2n_0-1}0^{m+1}1)$ because $f_0f_1<f_1f_0$.
  \end{proof}

  Now suppose $n_0=1$. We saw in the proof of \thref{claim2} above that
  $\rho(11)<\rho(00)$ implies $n_0=1$, but a priori both $\rho(11)<\rho(00)$ and
  $\rho(00)<\rho(11)$ are possible when $n_0=1$. Note that $\rho(10)$ is always
  minimal, however. The possible 3-minima are $\rho(1^3)$ (if $\rho(11)$ is
  maximal), $\rho(0^21)$ (if $\rho(00)$ is maximal), and $\rho(10^2)$ (in either
  case). But $\rho(1^3)> \rho(1^20)$ by $n_0=1$ and $\rho(0^21)>\rho(10^2)$
  because $f_1f_0^2 > f_0^2 f_1$, as observed in the base case of \thref{claim2}. Hence
  $\rho(10^2)$ is always the 3-minimum when $n_0=1$. We now split into two final
  subcases to finish the argument when $x_0>i_x$ and $n_0=1$. First, assume
  $\rho(00)< \rho(11)$, so $\rho(11)$ is maximal. For any $m\geq2$, if
  $\rho(10^{m-1}1)$ is maximal and $\rho(10^m)$ is minimal, the next maximum is
  $\rho(10^m1)$ since the other possibility is $\rho(10^{m-1}10)$, which is of
  the form $f_0f_1y$ for some $y$, and $f_0f_1y < f_1f_0y$. And in this case the
  next minimum is $\rho(10^{m+1})$, because the other option is
  $\rho(10^{m-1}1^2)$, which is of the form $f_1^2f_0y$ hence greater than
  $f_0f_1^2y$. So by induction $\rho(10^m1)$ is witnessed for all $m$ if
  $\rho(11)$ is maximal.

  The remaining subcase of $x_0>i_x$ is $n_0=1$ and $\rho(11)<\rho(00)$. In
  general, it may be that $\rho(0^{\ell-1})$ is maximal for finitely many $\ell
  \geq 3$, but this cannot be the case for all $\ell$ (as we assume $b<1$)
  because $\rho(0^\ell)$ decreases to $r_0$ as $\ell$ increases, while some
  probabilities of every length will be greater than $r_1$.
  Suppose $\rho(0^{\ell-1})$ is maximal and (by induction) $\rho(10^{\ell-2})$
  is minimal, for $\ell\geq3$. Then either $\rho(0^\ell)$ or
  $\rho(10^{\ell-2}1)$ is $\ell$-maximal, and $\rho(10^{\ell-1})$ is always
  $\ell$-minimal (by the same argument as in the last paragraph). If
  $\rho(10^{\ell-2}1)$ is maximal, then the argument in \thref{claim3} takes over from
  length $\ell+1$ onwards. If $\rho(0^\ell)$ is maximal, then $\rho(10^\ell)$ is
  $(\ell+1)$-minimal because the other option is $\rho(0^\ell1) >
  \rho(0^{\ell-1}10)$. The argument then repeats for $\ell+1$, and so on,
  meaning we witness $0^\ell$ for finitely many $\ell$ and then $10^m1$ for all
  large enough $m$.

  This completes the argument when $x_0>i_x$. If instead $x_0<i_x$, then
  something similar happens, but with even-odd parities switched. We state
  without detailed proofs the three claims (corresponding to those above) that
  will finish the argument here, as their proofs follow in the same way
  \textit{mutatis mutandis}. First, there is a least $n_0$ such that
  $\rho(1^{2n_0+2}) > \rho(1^{2n_0}0^2)$. The case $n_0=0$ is separate and
  exactly analogous to the case $n_0=1$ when $x_0>i_x$: if $n_0=0$ then
  $\rho(11) > \rho(00)$, so $\rho(00)$ is minimal, $\rho(01)$ is maximal, and in
  general we have $0^m$ minimal and $0^{m-1}1$ maximal for all $m\geq2$. So
  assume $n_0 > 0$ from now on.

  The second claim is that when $n_0>0$ and $x_0<i_x$, for all $1\leq \ell\leq
  n_0$, $\rho(1^{2\ell-2}01)$ is $2\ell$-maximal; $\rho(1^{2\ell})$ is $2
  \ell$-minimal; $\rho(1^{2\ell+1})$ is $(2\ell+1)$-maximal; and
  $\rho(1^{2\ell}0)$ is $(2\ell+1)$-minimal. Both the base case and the
  inductive step work very similarly as before (here, $\rho(11)$ being minimal
  relies on $n_0>0$). The only possible $(2\ell+2)$-maxima are $\rho(1^{ 2\ell}
  01)$ and $\rho(1^{2\ell+1}0)$; the only possible $(2\ell+2)$-minima are
  $\rho(1^{2\ell+2})$ and $\rho(1^{2\ell}0^2)$; the only possible $(2\ell+
  3)$-maxima are $\rho(1^{2\ell+3})$ and $\rho(1^{2\ell}010)$; and the only
  possible $(2\ell+3)$-minima are $\rho(1^{2\ell+2}0)$ and $\rho(1^{ 2\ell-2}
  01^2)$. All the alternatives listed can be dispensed with using
  $f_0f_1<f_1f_0$, $f_0f_1^2 < f_1^2f_0$, and $\ell<n_0$. (The latter is only
  needed to show the claim holds for $\ell+1$ given it holds for $\ell$, and so
  it does hold for $\ell= n_0$ as stated.)

  The third and last claim needed is that when $n_0>0$ and $x_0<i_x$, for all
  $m\geq1$, we have $\rho(1^{ 2n_0}0^m)$ minimal and $\rho(1^{2n_0}0^{m-1}1)$
  maximal. The case $m=1$ follows by taking $\ell=n_0$ in the previous claim.
  The inductive step is again very similar to \thref{claim3} for $x_0>i_x$, since the
  other possible minimum is $\rho(1^{2n_0}0^{m-1}1^2)$, which is of the form
  $f_1^2f_0y$ and hence not minimal since it is greater than $f_0f_1^2y$. The
  other possible maximum is $\rho(1^{2n_0}0^{m-1}10)$, of the form $f_0f_1y$,
  which is less than $f_1f_0y$ and hence not maximal.

  To finish the argument for Case 4(a), we dispense with the special case when $b=0$,
  i.e., when $f_0$ is constant. (The subcase where instead $f_1$ is constant was
  dealt with in Case 2.) Here $r_0=a$, and $\rho(w^\smallfrown 01) = f_1a$ for
  all strings $w$. If $x_0>i_x$, then proceeding as above, we see that after
  applying $f_1$ some number of times, if $n$ is least such that
  $\rho(1^{2n+1})\geq a$, then it is not possible for the probability of any
  string with length greater than $2n$ to exceed $f_1a$. This means that maximal
  probabilities cease to be unique at length $2n+1$, and only finitely many
  strings can be witnessed. The same holds when $x_0<i_x$, but now the maxima
  cease to be unique after $\rho(1^{2n})\geq a$.  

\end{itemize}

\noindent \textit{Strings witnessed in the above case:} $1^{2n-1}0^m1$,
$1^{2n}0^m1$, $0^m$, $0^m1$. 

\begin{itemize}[leftmargin=*]
  \item[(b)] Both $f_0$ and $f_1$ increase $i_x$. This is equivalent to $f_0f_1x >
    f_1f_0x$ for all $x$, so that appending $10$ always gives a higher
    probability than appending $01$. It is also equivalent to $f_0f_1^2x >
    f_1^2f_0x$ for all $x$ (see \thref{classobs}). As before, we put off the
    special case $b=0$ for later, and assume for the moment that $b>0$. 

    First, say that $x_0<i_x$. Here we need to split into slightly different
    subcases than we did in (a). Since $\rho(0)< \rho(1)$, the 2-maximum is
    always $\rho(10)$ because the only other option is $\rho(01) < \rho(10)$.
    The possible 2-minima are $\rho(00)$ and $\rho(11)$, and both
    $\rho(00)<\rho(11)$ and $\rho(11)<\rho(00)$ are possible. Suppose first that
    $\rho(00)<\rho(11)$. It may be that $\rho(0^\ell)$ is minimal for finitely
    many $\ell$, but eventually this is no longer the case since $\rho(0^\ell)$
    increases to $r_0$ while some other probabilities always stay below $r_1$.
    Suppose that for some $\ell\geq 2$, $\rho(0^\ell)$ is minimal and
    $\rho(10^{\ell-1})$ is maximal. The possible $(\ell+1)$-maxima are
    $\rho(10^\ell)$ and $\rho(0^\ell1)$, but the latter is of the form $f_1f_0y$
    for some $y$, which is less than $f_0f_1y$ and so not maximal. The possible
    $(\ell+1)$-minima are $\rho(0^{\ell+1})$ and $\rho(10^{\ell-1}1)$. Either
    may be the case in general, and if $\rho(0^{\ell+1})$ is minimal then the
    argument repeats for length $\ell+1$: now $\rho(10^\ell)$ is maximal. For
    large enough $\ell$, that is no longer the case, and for such an $\ell$ we
    have $\rho(10^{\ell-1})$ maximal and $\rho(10^{\ell-2}1)$ minimal. Once that
    happens, the $(\ell+1)$-maximum is $\rho(10^\ell)$ since
    $\rho(10^{\ell-2}1^2)$ is of the form $f_1^2f_0y$ for some $y$, which is
    less than $f_0f_1^2y$. The $(\ell+1)$-minimum is $\rho(10^{\ell-1}1)$ since
    the other option is $\rho(10^{\ell-2}10)$, which is of the form $f_0f_1y$,
    and this is greater than $f_1f_0y$. It follows by induction that we witness
    $10^m$ for all $m\geq0$ in this case.

    For the rest of the argument for $x_0<i_x$, we assume instead that
    $\rho(11)<\rho(00)$. The argument follows from a series of three claims,
    much like in part (a). First, there is a least $n_0$ such that
    $\rho(1^{2n_0-1}0^2) > \rho(1^{2n_0+1})$. Then $n_0\geq 1$. The case $n_0=1$
    requires special treatment, which we outline before proceeding further. We
    have $\rho(11)$ minimal and $\rho(10)$ maximal. Since $\rho(1^3)<\rho(10^2)$
    when $n_0=1$, the 3-maximum is $\rho(10^2)$, and the 3-minimum is
    $\rho(101)$ because the other option $\rho(1^20)$ is greater than
    $\rho(01^2)$. Inductively, if for $m\geq2$ we have $\rho(10^m)$ maximal and
    $\rho(10^{m-1}1)$ minimal, then the $(m+2)$-maximum is $\rho(10^{m+1})$
    since the other option, $\rho(10^{m-1}1^2)$, is of the form $f_1^2f_0y$,
    which is less than $f_0f_1^2y$ and so not maximal. And the $(m+2)$-minimum is
    $\rho(10^m1)$ since the other option is $\rho(10^{m-1}10)$, which is of the
    form $f_0f_1y$, which is greater than $f_1f_0y$ and so not minimal. It
    follows that we witness $10^m$ for all $m\geq0$ in this subcase.

    Now assume $n_0>1$ as well as $\rho(11)<\rho(00)$. The second claim to
    complete the proof is that for any $1\leq \ell \leq n_0$, we have that
    $\rho(1^{2\ell-1}0)$ is $2\ell$-maximal; $\rho(1^{2\ell})$ is
    $2\ell$-minimal; if $\ell<n_0$, then $\rho(1^{2\ell+1})$ is
    $(2\ell+1)$-maximal; and $\rho(1^{2\ell-1}01)$ is $(2\ell+1)$-minimal.  The
    induction argument goes exactly as in \thref{claim2} from case (a) where $x_0>i_x$,
    except switching the roles of ``maximal'' and ``minimal'' everywhere as well
    as switching the roles of (firstly) $f_0f_1$ and $f_1f_0$, and (secondly)
    $f_0f_1^2$ and $f_1^2f_0$. This is because we now have $f_1f_0<f_0f_1$ and
    $f_1^2 f_0 < f_0f_1^2$ by \thref{classobs}. The third claim, which completes
    the picture, is that $\rho(1^{2n_0-1}0^m)$ is maximal and
    $\rho(1^{2n_0-1}0^{m-1}1)$ is minimal for all $m\geq0$. The cases $m=0$ and
    $m=1$ follow from taking $\ell=n_0-1$ and $\ell=n_0$ in the second claim.
    For $m=2$, the $(2n_0+1)$-minimum is $\rho(1^{2n_0-1}01)$ by the second
    claim again, and the $(2n_0+1)$-maximum is $\rho(1^{2n_0-1}0^2)$ because the
    other option is $\rho(1^{2n_0+1})$, and this is the lesser value by
    definition of $n_0$. The induction can be carried out from here using
    $f_1^2f_0 < f_0f_1^2$ and $f_1f_0 < f_0f_1$, finishing the proof for
    $x_0<i_x$.

    Now suppose $x_0>i_x$. The proof of this case is split into three claims, as
    usual. First, there is a least $n_0$ such that $\rho(1^{2n_0+2}) \leq
    \rho(1^{2n_0} 0^2)$. As before, we first need to consider the case $n_0=0$
    separately, but fortunately this is equivalent to $\rho(11)<\rho(00)$ so
    there is no need for a third subcase as with the $x_0<i_x$ argument.  If
    $n_0=0$, then $\rho(00)$ is maximal since $\rho(1)< \rho(0)$, and $\rho(01)$
    is minimal by $f_1f_0<f_0f_1$. In general, suppose for any $m\geq2$ that
    $\rho(0^m)$ is $m$-maximal and $\rho(0^{m-1}1)$ is $m$-minimal. Then
    $\rho(0^{m+1})$ is $(m+1)$-maximal since the other option is
    $\rho(0^{m-1}1^2)$, which is of the form $f_1^2f_0y$, which is less than
    $f_0f_1^2y$ and so not maximal. And $\rho(0^m1)$ is $(m+1)$-minimal since
    the other option $\rho(0^{m-1}10)$ is of the form $f_0f_1y$, which is
    greater than $f_1f_0y$ and hence not minimal. It follows by induction that
    $0^m$ is witnessed in this subcase for all $m\geq1$.

    Assume from now on that instead $n_0>0$. The second claim we need to finish
    the proof is that for $1\leq \ell \leq n_0$, $\rho(1^{2\ell})$ is
    $2\ell$-maximal; $\rho(1^{2\ell-2}01)$ is $2\ell$-minimal;
    $\rho(1^{2\ell}0)$ is $(2\ell + 1)$-maximal; and $\rho(1^{2\ell+1})$ is
    $(2\ell+1)$-minimal. The base case here uses $n_0>0$ to show $\rho(11)$ is
    maximal. The third claim is that $\rho(1^{2n_0} 0^m)$ is maximal and
    $\rho(1^{2n_0}0^{m-1}1)$ is minimal for all $m\geq2$. Here, for $m=2$, we
    have $\rho(1^{2n_0}0^2)$ maximal since by definition of $n_0$,
    $\rho(1^{2n_0+2})$ cannot be. And $f_0f_1>f_1f_0$ implies that
    $\rho(1^{2n_0}01)$ is minimal rather than $\rho(1^{2n_0+1}0)$. The
    inductive steps of both claims can be shown in a straightforward way using
    $f_1f_0<f_0f_1$, $f_1^2f_0<f_0f_1^2$, and in the first statement of the
    second claim, $\ell<n_0$. (The latter is used only to show the second claim
    holds for $\ell+1$ given it holds for $\ell$, so it does hold for $\ell=n_0$
    as stated.)

    Finally, suppose $b=0$. As in Case 4(a), only finitely many strings can be
    witnessed. We have again that $r_0=a$ and $\rho(w^\smallfrown 01)=f_1a$ for
    all strings $w$. If $x_0<i_x$, and $n$ is large enough that
    $\rho(1^{2n+1})\leq a$, then no longer string can have probability greater
    than $f_1a$, and this value is never attained uniquely. If $x_0>i_x$, and
    $n$ is large enough that $\rho(1^{2n})\leq a$, the same conclusion holds.
    Therefore at most finitely many constant strings can be witnessed, and
    nothing else. This completes the proof of Case 4(b) and of
    \thref{classification2}.
\end{itemize}

\noindent \textit{Strings witnessed in the above case:} $1^{2n-1}0^m$,
$1^{2n}0^m$.

%% file: pictures-pos2.tex
\begin{tikzpicture} 
  \begin{axis}[grid=minor,xmin=0,xmax=1,ymin=0,ymax=1,
    axis lines=middle,
    domain=0:1,
    xtick={.57,.669,.806},
    ytick={0},
    xticklabels = {$r_0$,$r_1$,$i_x$},
    restrict y to domain=0:1]
    \addplot[black,samples=50,smooth,unbounded coords=discard]
    plot (\x, { .285 + .5*\x } )
    node [pos=0.85, above] {$f_0$};
    \addplot[black,samples=50,smooth,unbounded coords=discard]
    plot (\x, { .575 + .14*\x } )
    node [pos=0.9, below right] {$f_1$};
    \addplot[color=black,solid,mark=*] coordinates { (.57,.57) }; 
    \addplot[color=black,solid,mark=*] coordinates { (.669,.669) }; 
  \end{axis}
\end{tikzpicture}

%% file: pictures-pos3.tex
\begin{tikzpicture} 
  \begin{axis}[grid=minor,xmin=0,xmax=1,ymin=0,ymax=1,
    axis lines=middle,
    domain=0:1,
    xtick={.349,.577,.857},
    ytick={0},
    xticklabels = {$i_x$,$r_1$,$r_0$},
    restrict y to domain=0:1]
    \addplot[black,samples=50,smooth,unbounded coords=discard]
    plot (\x, { .3 + .65*\x } )
    node [pos=0.9, below] {$f_0$};
    \addplot[black,samples=50,smooth,unbounded coords=discard]
    plot (\x, { .45 + .22*\x } )
    node [pos=0.9, below right] {$f_1$};
    \addplot[color=black,solid,mark=*] coordinates { (.857,.857) }; 
    \addplot[color=black,solid,mark=*] coordinates { (.577,.577) }; 
  \end{axis}
\end{tikzpicture}

%% file: pictures-neg2.tex
\begin{tikzpicture} 
  \begin{axis}[grid=minor,xmin=0,xmax=1,ymin=0,ymax=1,
    axis lines=middle,
    domain=0:1,
    xtick={.48,.549,.792},
    ytick={0},
    xticklabels = {$r_0$,$r_1$,$i_x$},
    restrict y to domain=0:1]
    \addplot[black,samples=50,smooth,unbounded coords=discard]
    plot (\x, { .9-.64*\x } )
    node [pos=0.9, above, xshift=5pt, yshift=5pt] {$f_0$};
    \addplot[black,samples=50,smooth,unbounded coords=discard]
    plot (\x, { .615-.28*\x } )
    node [pos=0.9, below left] {$f_1$};
    \addplot[color=black,solid,mark=*] coordinates { (.549,.549) }; 
    \addplot[color=black,solid,mark=*] coordinates { (.48,.48) }; 
  \end{axis}
\end{tikzpicture}

%% file: pictures-neg3.tex
\begin{tikzpicture} 
  \begin{axis}[grid=minor,xmin=0,xmax=1,ymin=0,ymax=1,
    axis lines=middle,
    domain=0:1,
    xtick={.236,.427,.48},
    ytick={0},
    xticklabels = {$i_x$,$r_1$,$r_0$},
    restrict y to domain=0:1]
    \addplot[black,samples=50,smooth,unbounded coords=discard]
    plot (\x, { .7-.64*\x } )
    node [pos=0.9, below] {$f_1$};
    \addplot[black,samples=50,smooth,unbounded coords=discard]
    plot (\x, { .615-.28*\x } )
    node [pos=0.9, above right] {$f_0$};
    \addplot[color=black,solid,mark=*] coordinates { (.427,.427) }; 
    \addplot[color=black,solid,mark=*] coordinates { (.48,.48) }; 
  \end{axis}
\end{tikzpicture}

%% file: pictures-mix2.tex
\begin{tikzpicture} 
  \begin{axis}[grid=minor,xmin=0,xmax=1,ymin=0,ymax=1,
    axis lines=middle,
    domain=0:1,
    xtick={.425,.561,.769},
    ytick={0},
    xticklabels = {$r_0$,$r_1$,$i_x$},
    restrict y to domain=0:1]
    \addplot[black,samples=50,smooth,unbounded coords=discard]
    plot (\x, { .34+.2*\x } )
    node [pos=0.9, above, xshift=5pt, yshift=5pt] {$f_0$};
    \addplot[black,samples=50,smooth,unbounded coords=discard]
    plot (\x, { .74-.32*\x } )
    node [pos=0.9, below] {$f_1$};
    \addplot[color=black,solid,mark=*] coordinates { (.425,.425) }; 
    \addplot[color=black,solid,mark=*] coordinates { (.561,.561) }; 
  \end{axis}
\end{tikzpicture}

%% file: pictures-mix3.tex
\begin{tikzpicture} 
  \begin{axis}[grid=minor,xmin=0,xmax=1,ymin=0,ymax=1,
    axis lines=middle,
    domain=0:1,
    xtick={.177,.439,.589},
    ytick={0},
    xticklabels = {$i_x$,$r_1$,$r_0$},
    restrict y to domain=0:1]
    \addplot[black,samples=50,smooth,unbounded coords=discard]
    plot (\x, { .495+.16*\x } )
    node [pos=0.9, above] {$f_0$};
    \addplot[black,samples=50,smooth,unbounded coords=discard]
    plot (\x, { .58-.32*\x } )
    node [pos=0.9, above right] {$f_1$};
    \addplot[color=black,solid,mark=*] coordinates { (.589,.589) }; 
    \addplot[color=black,solid,mark=*] coordinates { (.439,.439) }; 
  \end{axis}
\end{tikzpicture}

%% file: classreverse.tex
\subsection{Proof of \thref{classification2-reverse}}\label{sec:classreverse}
We show that for every string $w$ listed in \thref{classification2-full}, there
is an IFS $(f_0,f_1,x_0)$ which falls into the subcase of the proof of
\thref{classification2} which would lead to $w$ being witnessed. This results in
a case breakdown into the following seven subfamilies of strings,
listed here with the subcases of \thref{classification2} which they employ:

\begin{itemize}
  \item $0^n1^m$ for a given $n$ and all $m$ -- Case 2(a), \thref{rev1};
  \item $1^{2n}0^m1$ for a given $n$ and all $m$ -- Case 4(a), \thref{rev6};
  \item $1^{2n-1}0^m1$ for a given $n$ and all $m$ -- Case 4(a), \thref{rev7};
  \item $1^{2n}(01)^m$ for a given $n$ and all $m$ -- Case 3(a), \thref{rev3};
  \item $1^{2n+1}(01)^m$ for a given $n$ and all $m$ -- Case 3(a), \thref{rev4};
  \item $1^{2n+1}0(10)^m$ for a given $n$ and all $m$ -- Case 3(b),
    \thref{rev5};
  \item $0^{2n}1(01)^m$ for a given $n$ and all $m$ -- Case 3(a),
    \thref{worstsubcase}.
\end{itemize}

The proofs all follow the same basic strategy, which goes roughly as follows.
Given $n$, derive an inequality equivalent to the condition from the relevant
subcase of the proof of \thref{classification2} which results in strings with
prefixes of length $n$ being witnessed. This translates to a requirement that
$x_0$ be chosen inside a certain interval depending on $n$ and the coefficients
of the IFS. For any fixed $n,a,b,c,d$, finitely many of these intervals will
overlap $[0,1]$, and the set of such intervals is closed downward in $n$:
for any $\ell\leq n$, if $n$'s interval overlaps $[0,1]$, so does $\ell$'s.
Derive an inequality $n< f(a,b,c,d)$ for some function $f$ which is equivalent
to $n$'s interval overlapping $[0,1]$.
Treat three out of $a,b,c,d$ as functions of the fourth and show that $f\to
\infty$ as the fourth number tends to $1$ or $-1$, depending on the subcase.
This finishes the proof since it shows that for infinitely many $n$, one can
choose $a,b,c,d$ to satisfy $n<f$, and this suffices.

Although all seven subcases follow this outline, the particularities are
different enough to warrant separate treatments, albeit with some details
omitted.

\begin{prop}\thlabel{rev1} $A_P(0^n1^m)=2$ for all $n,m\geq0$.
\end{prop}
\begin{proof} Let $n\geq1$ be given (the case $n=0$ is trivial). The IFS
  $(f_0,f_1,x_0)$ witnesses $0^n1^m$ for all $m$ if in Case 2(a) of the proof of
  \thref{classification2} with $x_0>i_x$, and if $n$ is least such that
  $f_0^nx_0 <i_x$, i.e., $f_0^nx_0 < i_x < f_0^{n-1}x_0$. Take $f_0=bx$ for
  $b<1$ and $f_1=c$ (so $a=d=0$). Then $f_0^nx_0 = b^nx_0$, and our condition
  becomes 
  \[
    b^nx_0 < i_x < b^{n-1}x_0 \quad\text{or equivalently}\quad
    \frac{i_x}{b^{n-1}} < x_0 < \frac{i_x}{b^n}.
  \]
  Since $b<1$, we have $i_x/b^n > i_x$ for all $n\geq1$. In order to choose
  $x_0$ to witness $0^n1^m$ for our given $n$, we need $i_x/b^{n-1} < 1$, or
  equivalently
  \[
    \frac{\log i_x}{\log b} + 1 > n.
  \]
  By increasing $b$ arbitrarily close to 1, and setting $c=b/2$ from $b$, we can
  make $\log i_x/\log b$ larger than any given $n$, so that it is possible to
  choose $x_0\in (i_x,1)$ in order for exactly $0^n1^m$ to be witnessed for all
  $m\geq0$.
\end{proof}

\begin{prop}\thlabel{rev6} $A_P(1^{2n}0^m1) = 2$ for all $n,m\geq0$.
\end{prop}
\begin{proof} Let $n \geq 1$ be given ($n=0$ is covered by the previous
  proposition). The IFS $(f_0,f_1,x_0)$ witnesses $1^{2n}0^m1$ for all $m\geq0$
  if in Case 4(a) of the proof of \thref{classification2}---that is, $b>0>d$ and
  both maps decrease $i_x$---if $x_0<i_x$, and if $n$ is least such that
  \[
    f_1^{2n+2}x_0 > f_0^2f_1^{2n}x_0.
  \]
  Thinking of the LHS here as $f_1^2f_1^{2n}x_0$, this inequality is equivalent
  to
  \begin{equation}\label{eq:F}\begin{gathered}
    a+ab + b^2f_1^{2n}x_0 < c+cd+d^2f_1^{2n}x_0 \quad\iff \quad
    F \coloneqq \frac{f_0^2 0 - f_1^2 0}{d^2-b^2} < f_1^{2n}x.
  \end{gathered}\end{equation}
  If $n$ is supposed to be the least number making $F<f_1^{2n}x_0$, then we
  would like $f_1^{2(n-1)}x_0 < F < f_1^{2n}x_0$. Note that for any $x$ and $n$,
  we have
  \[
    f_1^nx = c\sum_{i=0}^{n-1} d^i + d^nx = c\cdot \frac{1-d^n}{1-d} + d^nx =
    r_1(1-d^n) + d^nx,
  \]
  and an analogous formula for $f_0^nx$. Then
  \[
    f_1^{2(n-1)}x_0 < F \iff r_1(1-d^{2(n-1)}) + d^{2(n-1)}x_0 < F \iff x_0 <
    r_1 - \frac{r_1 - F}{d^{2(n-1)}},
  \]
  and on the other hand
  \[
    F < f_1^{2n}x_0 \iff x_0 > r_1 - \frac{r_1 - F}{d^{2n}}
  \]
  by a similar calculation. Now, in our situation it will always be the case
  that $F<r_1 = c/(1-d)$, because
  \begin{gather*}
    \frac{f_0^20 - f_1^20}{d^2-b^2} < \frac{c}{1-d} \iff a(1-d+b-bd) < c(1-b^2)
    \iff \frac{a}{1-b} < \frac{c}{1-d},
  \end{gather*}
  i.e., $r_0<r_1$. As long as we choose $a,b,c,d$ to make $r_0<r_1$, then, we
  have $F<r_1$. We also need $F>0$, but this will be guaranteed by
  $f_1^{2(n-1)}x_0 < F$ since the latter LHS is nonnegative for every $n\geq1$
  and $x_0$.

  Overall, then, the IFS witnesses $1^{2n}0^m1$ when we can pick $x_0$ such that
  \begin{equation}\label{eq:ivl6} 
    x_0 \in \paren{r_1 - \frac{r_1-F}{d^{2n}},r_1 - \frac{r_1-F}{d^{2(n-1)}}},
  \end{equation}
  a nonempty interval since $d^{2n}<d^{2(n-1)}$ and $r_1-F>0$. Both endpoints of
  this interval are less than $i_x$ if $r_0<r_1$, since $r_0<r_1$ iff $r_1<i_x$,
  and so choosing such an $x_0$ automatically fulfills the requirement that
  $x_0<i_x$. 

  We also need (for a given $n$) to be able to pick $x_0>0$, so at
  least the right endpoint in \eqref{eq:ivl6} should be positive. For any $n$.
  \begin{gather*}
    r_1 - \frac{r_1-F}{d^{2(n-1)}} > 0 \iff d^{2(n-1)} > 1- \frac{F}{r_1} \iff
      (n-1)\log d^2 > \log\paren{1- \frac{F}{r_1}}\\
    \iff n < 1 + \frac{\log(1 - F/r_1)}{\log d^2}.
  \end{gather*}
  For arbitrarily large $n$ to be possible, the last RHS must be able to grow
  arbitrarily large depending on $a,b,c,d$. To accomplish this we can treat $d$
  as a variable and make $c$ a function of $d$ (so that $F$ and $r_1$ are as
  well), then require that
  \begin{equation}\label{eq:lim6}
    \lim_{d\to-1^+} \frac{\log(1-F/r_1)}{\log d^2} = \infty.
  \end{equation}
  We need $c$ to be a function of $d$ because $c$ must be greater than $\abs{d}$
  for all $d > -1$ if we are to have $c+d>0$, so $c$ will necessarily approach
  $1$ in the limit. Of course we also need to make sure the logarithm in the
  numerator is defined for all $d>-1$. If so, then together with the fact that
  the right endpoint of \eqref{eq:ivl6} is always less than $i_x$, we will have
  that for \emph{every} $n\geq1$ there is a choice of $a,b,c,d,x_0$ making
  $(f_0,f_1,x_0)$ witness $1^{2n}0^m1$ for all $m$.

  So, to sum up thus far, we want to choose numbers $a,b$ and a continuous
  function $c(d)$ to satisfy the requirements that $\abs{d} < c(d) < 1$, $a
  < 1-b$, $r_0<r_1$, $i_x<1$, $i_y>0$, and the limit condition \eqref{eq:lim6}
  holds. Because we are taking the limit as $d\to-1^+$, we may as well only
  bother asking for the other requirements to hold in the limit, too. This
  simplifies things considerably: since $c\to1$ as $d\to -1$, we have $r_1\to
  1/2$. Then for $r_0<r_1$ to hold in the limit, it is enough to make
  $r_0=a/(1-b) < 1/2$, or in other words $2a < 1-b$. This condition also
  guarantees $a<1-b$ and hence $a+b\in [0,1]$. Furthermore, since $c(d)$ will
  eventually be greater than any fixed $a<1$, $a<c$ is satisfied in the limit.
  That $c+d\in [0,1]$ is implied by the requirement that $\abs{d}<c(d)<1$. 
 
  Only two conditions remain to be checked. Firstly, \eqref{eq:lim6} holds if
  $1-F/r_1$ stays strictly between $0$ and $1$ as $d\to-1^+$: on the one hand,
  $0< 1-F/r_1$ iff $F<r_1$, which as we saw is equivalent to $r_0<r_1$. On the
  other hand, $1 - F/r_1 < 1$ iff both $F$ and $r_1$ are positive, and both of
  those happen in the limit as noted above. Finally, we need to check that the
  lines intersect in $[0,1]^2$. But since $f_1(x) \to 1-x$ as $d\to-1$, if we
  make sure to take $a,b>0$, then $f_1$ will eventually intersect any line that
  stays inside $[0,1]^2$. Hence $i_y>0$ and $i_x<1$ hold in the limit, and we
  are done.
\end{proof}

The proofs of all but one of the remaining cases are very similar to the above,
and we will give a somewhat more streamlined presentation from here on out. The
most complicated subcase we save for last (\thref{worstsubcase}).

\begin{prop}\thlabel{rev7} $A_P(1^{2n-1}0^m1) = 2$ for all $n \geq 1$, $m\geq0$.
\end{prop}
\begin{proof} If $n \geq 1$ is given, then $(f_0,f_1,x_0)$ witnesses
  $1^{2n-1}0^m1$ for all $m\geq0$ if in Case 4(a) of the proof of
  \thref{classification2} (mixed slopes) with $x_0>i_x$ and $n$ least such that
  \begin{equation}\label{eq:case7-1}
    f_1^{2n+1}x_0 > f_0^2f_1^{2n-1}x_0.
  \end{equation}
  As before, we will pick $a,b>0$ constants and $c$ a continuous function of $d$
  so that as $d\to-1^+$, it is always possible to choose $x_0$ making the above
  happen for any given $n$. Now, \eqref{eq:case7-1} is equivalent to
  \begin{equation}\label{eq:case7-1a}
    f_1^{2n-3}x_0 < E < f_1^{2n-1}x_0 \quad\text{where}\quad E = \frac{a(1+b) -
    c(1+d)}{d^2-b^2},
  \end{equation}
  and the inequality in \eqref{eq:case7-1a} is equivalent to
  \begin{equation}\label{eq:7-2}
    x_0 \in \paren{r_1 + \frac{r_1-E}{\abs{d}^{2n-3}}, r_1 +
    \frac{r_1-E}{\abs{d}^{2n-1}}}.
  \end{equation}
  For arbitrarily large $n$ to be possible, we want to pick $a,b,c,d$ so that
  this interval intersects $(i_x,1)$, so a suitable $x_0$ can be chosen. The
  left endpoint in \eqref{eq:7-2} can be made less than $1$ for arbitrarily
  large $n$ if, in particular,
  \begin{equation}\label{eq:lim7}
    \lim_{d\to -1^+} \frac{\log \dfrac{r_1-E}{1-r_1}}{\log d^2} + \frac{3}{2} =
    \infty.
  \end{equation}
  And the right endpoint in \eqref{eq:7-2} is greater than $i_x$, for a
  given $n$, iff
  \begin{equation}\label{eq:7-3}
    \frac{\log \dfrac{r_1-E}{i_x-r_1}}{\log d^2} + \frac{1}{2} > n.
  \end{equation}
  Note that in the limit, $E$ approaches $r_0$ (as long as $b<1$). Hence as long
  as $r_0<r_1$ in the limit, then eventually $r_1>E$. If we arrange things so
  $i_x$ stays below $1$, then, 
  \[
    \frac{r_1-E}{i_x-r_1} > \frac{r_1-E}{1-r_1}.
  \]
  Also notice that we can take $\frac{r_1-E}{i_x-r_1}<1$ in the limit since this
  is equivalent to $2r_1<i_x+E$, which in the limit is guaranteed if $2a+b<1$,
  as may be checked with a little algebra. Assume that $a,b$ are positive with
  $2a+b<1$. Then since $\log$ is increasing, if \eqref{eq:lim7} holds, the LHS
  of \eqref{eq:7-3} will also approach $\infty$. This implies that whenever $n$
  is such that the left endpoint of \eqref{eq:7-2} is less than $1$, for all
  $n'\leq n$ it is possible to choose $x_0\in (i_x,1)$ in order to witness
  $1^{2n'-1}0^m1$.

  So, let $c(d)$ be a continuous function with $\abs{d}<c(d)<1$ for all $d>-1$,
  and let $a,b>0$ be such that $2a+b<1$. This immediately implies $a+b,c+d \in
  [0,1]$ for all $d$. Since $r_1\to 1/2$ as $d\to-1$, we have $r_0<r_1$ in the
  limit since $r_0 = a/(1-b) < 1/2$. We also need $E>0$, which is guaranteed as
  $d\to-1$ since $E$ approaches $r_0>0$. Since $c\to1$ and $d\to-1$, eventually
  $c>a$ as required. Because $f_1(x) \to 1-x$ as $d\to-1$, $c+dx$ will
  eventually intersect $a+bx$ in $[0,1]^2$, so that $0<i_y<i_x<1$. It only
  remains to check \eqref{eq:lim7}. But we already observed that
  \[
    \frac{r_1-E}{1-r_1} < \frac{r_1-E}{i_x-r_1} < 1
  \]
  as $d\to-1$, and $\frac{r_1-E}{1-r_1}>0$ iff $r_1>E$, which also holds in the
  limit. Therefore the logarithm in the numerator approaches a finite negative
  number, while $\log d^2$ approaches $0$ from below.
\end{proof}

\begin{prop}\thlabel{rev3} $A_P(1^{2n}(01)^m) = 2$ for all $n,m\geq0$.
\end{prop}
\begin{proof} For a given $n$, we witness $1^{2n}(01)^m$ if in Case 3(a) of
  the proof of \thref{classification2} (both slopes negative), with $x_0>i_x$
  and $n$ least such that
  \begin{equation}\label{eq:3-1}
    f_1^{2n}x_0 < i_x.
  \end{equation}
  We will pick numbers $a>0$, $b<0$, and a continuous function $c(d)$ so that as
  $d\to-1^+$, we have $a+b,c+d\in [0,1]$, $a<c$, $b>d$, $r_0<r_1$, and the lines
  $a+bx$ and $c+dx$ intersecting in $[0,1]^2$. If $a,b\notin \{0,\pm1\}$, then
  the last condition is automatically met as $d\to-1$ since $f_1\to 1-x$ and
  this intersects any line in $[0,1]^2$. The conditions $a<c$ and $b>d$ are also
  automatically met as $d\to-1$. At the same time, we must (given $n$) be able
  to pick
  \begin{equation}\label{eq:ivl3}
    x_0 \in \paren{ r_1+ \frac{i_x-r_1}{d^{2(n-1)}}, r_1 +
    \frac{i_x-r_1}{d^{2n}}}
  \end{equation}
  so that $f_1^{2n}x_0 < i_x < f_1^{2(n-1)}x_0$. We need this interval to
  intersect $(i_x,1)$ for arbitrarily large $n$, for suitable choices of
  $a,b,c,d$. That the right endpoint is always greater than $i_x$, for any $n$,
  follows from $d^{2n}<1$, since then $\frac{i_x-r_1}{d^{2n}} > i_x-r_1$. For
  the left endpoint to be less than $1$ for arbitrarily large $n$ we need
  \begin{equation}\label{eq:lim3}
    \lim_{d\to-1^+} \frac{\log \dfrac{i_x-r_1}{1-r_1}}{\log d^2} = \infty.
  \end{equation}

  Pick $a>0>b$ with
  \begin{equation}\label{eq:3-2}
    -b < a < \frac{1-b}{2}.
  \end{equation}
  Also let $c(d)$ be a continuous function with $\abs{d} < c(d) < 1$ for all
  $d>-1$. This immediately gives $c+d\in [0,1]$, and \eqref{eq:3-2} implies $a+b
  \in [0,1]$ too. Next, since $r_1\to 1/2$ as $d\to-1$ and \eqref{eq:3-2} makes
  $r_0 = a/(1-b) < 1/2$, we have $r_0<r_1$ in the limit. Finally, to satisfy
  \eqref{eq:lim3}, we want $\frac{i_x-r_1}{1-r_1}$ to be strictly between $0$
  and $1$ in the limit. This quantity is automatically positive since $i_x>r_1$
  and $1>r_1$ (both in the limit, again). And because \eqref{eq:3-2} implies
  $i_x \to \frac{1-a}{b+1} < 1$ as $d\to-1$, the fraction is also less than $1$
  in the limit. This completes the proof.
\end{proof}

\begin{prop}\thlabel{rev4} $A_P(1^{2n+1}(01)^m) = 2$ for all $n,m\geq0$.
\end{prop}
\begin{proof} Given $n$, take the IFS to be in Case 3(a) of
  the proof of \thref{classification2} (both slopes negative) with $x_0 < r_0$ and $n$ such
  that
  \begin{equation}\label{eq:4-1}
    f_1^{2n+1}x_0 < i_x < f_1^{2n-1}x_0.
  \end{equation}
  This is equivalent to
  \begin{equation}\label{eq:ivl4}
    x_0 \in \paren{ r_1 - \frac{i_x-r_1}{\abs{d}^{2n+1}}, r_1 - \frac{i_x-r_1}{
    \abs{d}^{2n-1}}}.
  \end{equation}
  Since
  \[
    r_1 - \frac{i_x-r_1}{\abs{d}^{2n+1}} < r_0 \iff \abs{d}^{2n+1} <
    \frac{i_x-r_1}{r_1-r_0}
  \]
  and the last fraction is greater than $1$ by \thref{classobs}(e) while the LHS
  is less than $1$, we have that the left endpoint of \eqref{eq:ivl4} is always
  less than $r_0$ for all $n\geq0$. In order to make the right endpoint of
  \eqref{eq:ivl4} greater than $0$ for arbitrarily large $n$ (for suitable
  choice of $a,b,c,d$), so that an $x_0 \in (0,r_0)$ may be chosen to make the
  IFS witness exactly $1^{2n+1}(01)^m$, we can arrange for
  \begin{equation}\label{eq:lim4}
    \lim_{d\to-1^+} \frac{ \log(i_x/r_1 - 1)}{\log d^2} = \infty.
  \end{equation}

  As usual, pick constants $a,b\notin \{0,\pm1\}$, $a>0>b$, and a continuous
  function $c(d)$ such that $\abs{d} < c(d)<1$ for all $d>-1$ (so $c+d\in
  [0,1]$). To satisfy \eqref{eq:lim4}, we want $0 < i_x/r_1 - 1<1$ in the limit,
  or equivalently $r_1 < i_x < 2r_1$. Since $i_x$ converges to $(1-a)/(b+1)$ and
  $r_1\to 1/2$, this can achieved (along with $a+b\in [0,1]$) by making $-b < a<
  \frac{1-b}{2}$. This implies that $a<c$ and $b>d$ are met in the limit, and
  again since $f_1 \to 1-x$ we will eventually have $(i_x,i_y)\in [0,1]^2$. And
  $r_0<r_1$ follows from $r_1<i_x$.
\end{proof}

\begin{prop}\thlabel{rev5} $A_P(1^{2n+1}0(10)^m) = 2$ for all $n,m\geq0$.
\end{prop}
\begin{proof} For this, given $n$, we take the IFS to be in Case 3(b) of
  the proof of \thref{classification2}, so that both maps have negative slope and
  \emph{increase} $i_x$. We want $x_0>r_0$ and $n$ to be such that
  \begin{equation*}\label{eq:5-1}
    f_1^{2n-1}x_0 < i_x < f_1^{2n+1}x_0.
  \end{equation*}
  This is equivalent to
  \begin{equation}\label{eq:ivl5}
    x_0 \in \paren{r_1 + \frac{r_1-i_x}{\abs{d}^{2n-1}}, r_1 + \frac{r_1-i_x}{
    \abs{d}^{2n+1}}}.
  \end{equation}
  Remember that in the present case we have $i_x<r_1<r_0$. We will pick $a>0>b$
  with
  \begin{equation}\label{eq:5-ab}
    \frac{1-b}{2} < a< 1,
  \end{equation}
  and pick $c(d)$ a continuous function with $\abs{d}<c(d)<1$ for all $d>-1$.
  Then if we take $d\to-1$, we have $r_1\to 1/2$ and $r_0 >1/2$ by choice of $a$
  and $b$, so that $r_1<r_0$ in the limit. Also $a+b,c+d\in [0,1]$, $a<c$, and
  $b>d$ hold in the limit; and as before, $a+bx$ eventually intersects $c+dx$ in
  $[0,1]^2$ since $c+dx \to 1-x$. Now we just need to make sure we can always
  pick an $x_0 \in (r_0,1)$ for arbitrarily large $n$ as $d\to-1$. We have
  \[
    r_1 + \frac{r_1-i_x}{\abs{d}^{2n+1}} > r_0 \iff \frac{r_1-i_x}{r_0-r_1} >
    \abs{d}^{2n+1}.
  \]
  Since the RHS here is less than $1$ and the LHS is greater than $1$ (by
  \thref{classobs}(e) again), this always happens for any $n$. To make the left
  endpoint of \eqref{eq:ivl5} less than $1$ for any given $n$, so that
  suitable $a,b,c,d,x_0$ may be chosen to witness the desired string, it
  suffices to ensure that
  \begin{equation}\label{eq:lim5}
    \lim_{d\to-1^+} \frac{\log \dfrac{r_1-i_x}{1-r_1}}{\log d^2} = \infty.
  \end{equation}
  Thus we want $0 < \frac{r_1-i_x}{1-r_1} < 1$ in the limit, or equivalently
  $2r_1-1 < i_x < r_1$. Since $r_1\to 1/2$, in the limit the latter inequality becomes
  \[
    0 < \frac{1-a}{b+1} < \frac{1}{2},
  \]
  which is equivalent to \eqref{eq:5-ab}.
\end{proof}

Now we arrive at the final and most complex subcase of
\thref{classification2-reverse} to prove. The extra difficulty arises because,
basically, we will need to take both $b$ and $d$ to $-1$ while both $a$ and $c$
go to $1$. This makes it harder to make certain properties hold ``in the limit''
as in the previous subcases, and also results in a limit condition in which the
limit converges to $\log \frac{0}{0}$. Slightly more delicate handling is needed
to get around these problems.

\begin{prop}\thlabel{worstsubcase} $A_P(0^{2n}1(01)^m) = 2$ for all $n,m\geq0$.
\end{prop}
\begin{proof} Let $n$ be given. The IFS $(f_0,f_1,x_0)$ witnesses
  $0^{2n}1(01)^m$ for all $m$ if in Case 3(a) of the proof of
  \thref{classification2} (where both maps have negative slope and both decrease
  $i_x$), when $x_0 > i_x$ and when $n$ is least such that $f_0^{2n}x_0<i_x$,
  i.e., 
  \[
    f_0^{2n}x_0 < i_x < f_0^{2(n-1)}x_0,
  \]
  or equivalently (after rearranging)
  \begin{equation}\label{eq:ivl2}
    x_0 \in \paren{r_0 + \dfrac{i_x-r_0}{b^{2n-2}}, r_0 +
    \dfrac{i_x-r_0}{b^{2n}}}.
  \end{equation}
  If we can pick $a,b,c,d$ to make $r_0<i_x$, then this interval is nonempty
  with positive endpoints. For this $n$ and $a,b,c,d$, it is possible to choose
  $x_0$ to witness the desired family of strings iff the left endpoint is less
  than $1$ and the right endpoint is greater than $i_x$. First,
  \[
    r_0 + \frac{i_x-r_0}{b^{2n}} > i_x \iff 1 > b^{2n},
  \]
  which is true for all $n\geq1$, so if an $x_0$ can be chosen above $i_x$ for a
  given $n$ then a suitable $x_0$ can also be chosen for any $n'\leq n$. And we
  can choose $x_0<1$ iff
  \begin{equation*}
    r_0 + \dfrac{i_x-r_0}{b^{2n-2}} < 1 \iff \dfrac{i_x-r_0}{1-r_0} < b^{2n-2}
    \iff \dfrac{\log \dfrac{i_x-r_0}{1-r_0}}{\log b^2} + 1 > n.
  \end{equation*}
  This is possible to achieve for any given $n$ if we can make
  \begin{equation}\label{eq:limcond2}
    \lim_{b\to -1^+} \dfrac{\log \dfrac{i_x-r_0}{1-r_0}}{\log b^2} =
    \infty.
  \end{equation}
  Altogether this means that if $r_0 + (i_x-r_0)/b^{2n} < 1$ for some $n$ and
  a fixed choice of $a,b,c,d$, then it is possible for every $1\leq n'\leq n$ to
  pick a suitable value of $x_0 > i_x$ making $(f_0,f_1,x_0)$ witness the
  strings $0^{2n'}1(01)^m$ for every $m \geq 0$. Hence the proof will be complete
  if we can choose $a$, $c$, and $d$ as functions of $b$ such that such that
  \eqref{eq:limcond2} holds and such that the IFS remains in Case 3(a) of the
  proof of \thref{classification2} for all $b>-1$. Actually, for technical
  reasons it will be simpler for now to choose $r_0$ as a function of $b$ and
  then let $a(b) = (1-b)r_0(b)$. This is not a problem because $b$ is never $1$,
  so $r_0(b) = a(b)/(1-b)$ is always well-defined. 
  We will ultimately see that the requirements we impose on $r_0(b)$ do not
  contradict the behavior of $a(b)$.

  We proceed by deriving necessary conditions on $r_0,c,d$ to satisfy each
  requirement, and showing along the way that each new condition is compatible
  with all the preceding ones. This will imply that functions $r_0,c,d$
  satisfying all of them do indeed exist. Our first requirements, which we will
  take as ``atomic'' in that they will not reduce to other requirements, are
  that
  \begin{equation}\label{eq:atomic}
    (1-b)r_0(b)<1 \quad\text{and}\quad \abs{b} < \abs{d(b)} < c(b) < 1
  \end{equation}
  for all $b>-1$ (with $b,d$ negative). The second of these immediately implies
  $c+d\in [0,1]$. To guarantee $a+b\in [0,1]$, first note that $a+b=r_0(1-b)+b <
  1$ iff $r_0<1$, and this follows from the first atomic requirement. Then $a+b
  > 0$ iff 
  \begin{equation}\label{eq:aplusbpos}
    r_0 > -b/(1-b),
  \end{equation}
  a new requirement. Actually, \eqref{eq:aplusbpos} will turn out to be a
  consequence of $i_x,i_y \in [0,1]$, or in other words of $f_0$ and $f_1$
  intersecting in $[0,1]^2$. We need the latter to happen anyway, so let us now
  find a sufficient condition for it.
  Rewriting $i_x$ and $i_y$ in terms of $r_0$ produces
  \[
    i_x = \dfrac{c-r_0(1-b)}{b-d} \quad\text{and}\quad i_y
    =\dfrac{bc-r_0(1-b)d}{b-d}.
  \]
  If $i_y<i_x$, or equivalently $r_0<r_1$, then it suffices to make $i_y>0$ and
  $i_x<1$.
  We will see how to ensure $r_0<r_1$ in a moment. One can check that
  \begin{equation}\label{eq:2-ixiy}
    i_y > 0 \iff r_0 > \dfrac{bc}{d(1-b)} \quad\text{and}\quad i_x<1 \iff
    r_0 > \dfrac{c+d-b}{1-b}.
  \end{equation}
  Since $c+d>0$, we have $\frac{c+d-b}{1-b} > \frac{-b}{1-b}$, so that
  satisfying \eqref{eq:2-ixiy} would automatically result in
  \eqref{eq:aplusbpos} being satisfied too. Thus \eqref{eq:aplusbpos} is
  redundant. Next, some more algebra shows that
  \[
    \frac{bc}{d(1-b)} < \frac{c+d-b}{1-b} \iff b>d,
  \]
  an atomic requirement. Hence the first condition in \eqref{eq:2-ixiy} is
  implied by the second as long as \eqref{eq:atomic} holds, so is also
  redundant.
  Then we will have $a+b>0$, $i_y>0$, and $i_x<1$ if we can choose $r_0$ so that
  \begin{equation}\label{eq:alphaivl}
    \frac{c+d-b}{1-b} < r_0 < \frac{c}{1-d} = r_1.
  \end{equation}
  The latter guarantees that $i_y<i_x$ so that we stay in Case~3(a) of the proof
  of \thref{classification2}, and also subsumes the second condition in
  \eqref{eq:2-ixiy}, so if \eqref{eq:alphaivl} holds then \eqref{eq:2-ixiy} is
  fully redundant.
  Now, the interval in \eqref{eq:alphaivl} is nonempty because
  \[
    \frac{c+d-b}{1-b} < \frac{c}{1-d} \iff (c+d-b)(1-d) < c(1-b) \iff
    (c-1+d)(b-d) < 0,
  \]
  which follows from the second requirement in \eqref{eq:atomic}: $b-d>0$ since
  $b>d$, and $c-1+d<0$ since $\abs{d}<c<1$. So \eqref{eq:atomic} makes it
  possible to choose $r_0$ to satisfy \eqref{eq:alphaivl}, and together
  \eqref{eq:atomic} and \eqref{eq:alphaivl} are enough to ensure we stay in
  Case~3(a).

  It remains to show that the limit requirement \eqref{eq:limcond2} is
  consistent with \eqref{eq:atomic} and \eqref{eq:alphaivl}. We will take $r_0$,
  $c$, and $d$ to be continuously differentiable functions of $b$. $\log b^2$
  approaches $0$ from below as $b\to -1^+$, so in order for the limit to reach
  $+\infty$, one needs the logarithm in the numerator to stay negative. For
  this, one wants to maintain
  \[
    0< \frac{i_x-r_0}{1-r_0} < 1
  \]
  in the limit, and for this quantity to stay strictly below $1$ at $b=-1$. Now,
  $d(b) \to -1^+$ as $b\to -1^+$ since $d$ is always less than $b$, and $c(b)
  \to 1$. Then after yet more algebra, we have that
  \[
    \frac{i_x-r_0}{1-r_0} = \frac{c - r_0(1-d)}{(b-d)(1-r_0)} \to
    \frac{0}{0} \quad\text{as } b\to-1.
  \]
  An application of L'H\^opital's Rule shows that the limit is equal to
  \begin{equation}\label{eq:2lim}
    \lim_{b\to-1^+} \frac{c' - r_0'(1-d)+r_0 d'}{(1-r_0)(1-d') -
    r_0' (b-d)} = \frac{2c'(-1) - 4r_0'(-1) + d'(-1)}{1-d'(-1)}.
  \end{equation}
  (The calculation follows since $r_0'$, $c'$, and $d'$ are bounded
  everywhere by assumption, and $r_0 \to 1/2$.) Since $d$ decreases to $-1$
  as $b$ decreases to $-1$, $d'(-1)\geq 0$, and we will need $d'(-1)\neq 1$ for
  \eqref{eq:2lim} to be well-defined. If
  we take $0<d'(-1)<1$, then the denominator of the limit in \eqref{eq:2lim} is
  positive. Hence the limit in \eqref{eq:limcond2} will tend to
  $\dfrac{-\infty}{0^-} = + \infty$, as needed, if
  \begin{equation}\label{eq:2prefinal}
    0< \frac{2c'(-1) - 4r_0'(-1) + d'(-1)}{1-d'(-1)} < 1.
  \end{equation}

  If $L(b) = \frac{c+d-b}{1-b}$ is the lower bound in \eqref{eq:alphaivl}, then
  one can calculate
  \begin{gather*}
    L'(-1) = \frac{2c'(-1) + 2d'(-1)-1}{4}, \quad r_0'(-1) =
    \frac{2a'(-1)+1}{4},\\
    \text{and} \quad r_1'(-1) = \frac{2c'(-1) + d'(-1)}{4}.
  \end{gather*}
  Using these expressions we see that \eqref{eq:2prefinal} is equivalent to
  \begin{equation}\label{eq:2final}
    L'(-1) < r_0'(-1) < r_1'(-1).
  \end{equation}
 
  Our final objective is to show \eqref{eq:2final} is consistent with the other
  requirements \eqref{eq:atomic} and \eqref{eq:alphaivl}, which will complete
  the proof since that means \eqref{eq:limcond2}, \eqref{eq:atomic}, and
  \eqref{eq:alphaivl} can all be satisfied simultaneously.
  Actually, under the above assumption that $0<d'(-1)<1$, and up to possibly
  perturbing $r_0$, $c$, and $d$, \eqref{eq:2final} is equivalent to
  \eqref{eq:alphaivl} holding in the limit. This follows because for any
  continuously differentiable functions $f(x),g(x)$ having the same limit as
  $x\to C^+$, where $C$ is some constant, then for any $\varepsilon>0$,
  $f(x)>g(x)$ on $(C,C+\varepsilon)$ iff $f'(x)>g'(x)$ on $(C,C+\varepsilon)$.
  Then since $L$, $r_0$, and $r_1$ all tend to $1/2$ as $b\to -1$, we have that
  \eqref{eq:alphaivl} holding in a right neighborhood of $b=-1$ is equivalent to
  $L' < r_0' < r_1'$ holding in the same neighborhood. By smoothly perturbing
  $r_0$, $c$, and $d$ if necessary, as long as $0<d'(-1)<1$ is maintained, we
  can ensure strict inequality between the derivatives holds at $b=-1$, i.e.,
  that \eqref{eq:2final} holds. (A bit more formally, one could say that these
  strict inequalities are all open conditions in the $C^1$ topology.) Thus
  \eqref{eq:2final} implies \eqref{eq:alphaivl} holds near $b=-1$, and
  conversely, \eqref{eq:alphaivl} implies that $r_0$, $c$, and $d$ may be taken
  to satisfy \eqref{eq:2final} and hence \eqref{eq:limcond2}. In particular,
  \eqref{eq:2final} and \eqref{eq:atomic} are also consistent with each other.

  So to sum up, there are continuously differentiable functions $r_0(b)$, $c(b)$,
  and $d(b)$ (and consequently $a(b) = (1-b)r_0(b)$) satisfying
  \eqref{eq:atomic}, \eqref{eq:alphaivl}, and $0<d'(-1)<1$. We have established
  that all of this suffices to be able to choose, given any $n$, values of $x_0$
  and $b$ which result in the IFS $(f_0,f_1,x_0)$ witnessing the strings
  $0^{2n}1 (01)^m$ for all $m\geq 0$.
  This finishes the proof of the final
  subcase of \thref{classification2-reverse}, and at last the proof of
  \thref{classification2-full} is complete.
\end{proof}

%% file: classafter.tex
\subsection{Further remarks}\label{sec:classafter}
The proof of \thref{classification2-reverse} appears to explicitly rely on the
use of IFSs over a two-letter alphabet, and a priori does not extend to show
that, e.g., $A_P(0^n1^n)=2$ may be witnessed by an IFS over $\{0,1,2\}$, for
which another map $f_2$ must be specified.
However, if one defines $f_0x = a+bx$ and $f_1x = c+dx$ as in any of the proofs
in the last section, and lets $f_jx = \frac{a+c}{2} + \frac{b+d}{2}x$ for all
other $j\in \Sigma$, then $f_jx$ is strictly between $f_0x$ and $f_1x$ except at
$x = i_x$, and so a string containing a $j$ can have neither minimal nor maximal
probability. Hence $A_P(w)=2$ over a two-letter $\Sigma$ implies that $A_P(w)=2$
over any $\Sigma' \supset \Sigma$. However, adding more maps to the IFS
certainly seems like it should generally decrease gaps, which leads us to

\begin{conj}
  $A_P$ is alphabet-independent, but $A_{P,\delta}$ is not for any $\delta>0$.
\end{conj}

\thref{classification2-full} immediately implies that the set of binary strings
with $A_P=2$ is a regular language. More particularly, the proof of
\thref{classification2} has the following consequence, which is somewhat
intriguing given that stochastic languages---which are defined by fixed
probability thresholds (the cut-point)---are not generally regular, or even
recursively enumerable, although Rabin did show that a stochastic language
defined by an isolated cut-point is regular \cite{R63}.
\begin{cor} For every two-state PFA $M$ over a binary alphabet, the language of
  strings whose complexity is witnessed by $M$ is regular.
\end{cor}
\begin{proof} By the proof of \thref{classification2}, given $M$, the set of
  strings $W$ witnessed by $M$ consists of one of the following plus at most
  finitely many other strings:
  \begin{itemize}
    \item nothing,
    \item $0^n$ for all $n$,
    \item $0^n1^m$ for some $n$ and all $m$,
    \item $0^n(10)^m$ for some $n$ and all $m$,
    \item $0^n1(01)^m$ for some $n$ and all $m$,
    \item $1^n(01)^m$ for some $n$ and all $m$,
    \item $1^n0(10)^m$ for some $n$ and all $m$,
    \item $1^n0^m1$ for some $n$ and all $m$, or
    \item the set of bit-flips of any one of the above.
  \end{itemize}
  In all cases, for all but finitely many $w$, we have $w\in W$ if and only if
  $w$ begins with a fixed prefix and ends with a repeated pattern of length 1 or
  2, possibly followed by a single extra digit. Each case can be described
  by a regular expression.
\end{proof}

Another consequence of the classification is that we can 
save an arbitrarily high number of states by switching from NFAs to PFAs to
describe a given (binary) string:

\ANAPdiff
\begin{proof}
  The statement follows if we can show $A_N(0^n1^n)$ is unbounded in
  $n$,\footnote{\cite[Theorem~12]{SW01} establishes that
    $A_D(0^n1^n) \geq \sqrt{n} -1$ for all $n$, but the proof does not quite
    go through for NFAs. Probably a similar explicit lower bound on $A_N$ can be
  found.} since $A_P(0^n1^n) = 2$ for all $n$ by \thref{classification2-full}.
  Suppose $A_N(0^n1^n) \leq K$ for all $n$ and some constant $K$. For any $w$,
  $A_N(w)$ can be witnessed by an NFA whose unique accepting path of length
  $\abs{w}$ uses every edge. Hence by the pigeonhole principle, there is some
  NFA $M$ with at most $K$ states such that for infinitely many $n$, there is a
  unique path of length $2n$ which accepts $0^n1^n$ and uses every edge of $M$.
  We show this is impossible. First, if the digraph of $M$ has fewer than two
  distinct cycles, then at most one string of the form $0^n1^n$ is accepted.
  Then we can assume there are distinct cycles of lengths $a$ and $b$,
  respectively. 
  For any string
  $w$ accepted by $M$, the portion of $w$ which was read while traversing these
  cycles has length $\ell = ax+by$ for some $x,y\in \N$. If such an $\ell$
  is greater than $2ab - a - b$, then there are at least two different pairs of
  natural numbers $(x,y)$ and $(x',y')$ with $ax + by = ax' + by' = \ell$ (see,
  e.g., \cite[Lemma~11]{SW01}). In terms of $M$, this means for all large enough
  $m$ such that $M$ accepts a word of length $m$ with a path that uses both
  cycles, there are at least two distinct accepting paths of length
  $m$---corresponding to traversing the cycles $x$ and $y$ times on the one
  hand, and $x'$ and $y'$ times on the other.
  In particular, the accepting path for $0^n1^n$ uses both cycles for infinitely
  many $n$ such that $A_N(0^n1^n)$ is witnessed by $M$, and so for all but
  finitely many of these $n$ there are two different accepting paths of length
  $2n$, a contradiction.
\end{proof}

Of course, the 2-state PFA describing $0^n1^n$ may have to be somewhat
complicated, a problem we briefly return to in Section~\ref{sec:otherdefs} below.

As remarked earlier, no evidence has yet appeared to suggest that $A_P$ is
unbounded, or even that any string has complexity greater than $3$. All binary
strings of length $9$ or less have complexity either $2$ or $3$, and witnesses
with three states have been found for a number of longer strings as well.
Therefore, we may pose the following questions, the first being restated from
the introduction:

\unbndq*
\begin{que} What is a tight upper bound for $A_P(w)$ as a function of $\abs{w}$?
\end{que}

Lastly, one may call a string \bdef{random} for a measure of complexity if its
complexity is the maximum possible for its length. For example, a string is
random for Kolmogorov complexity if its complexity is equal to its length, up to
an additive constant not depending on the string. 
For $A_N$, the string $w$ is random if $A_N(w) = \floor{\abs{w}/2}+1$, and this
is known to be tight (except over a binary alphabet; see
\cite[Theorem~9]{KComplexity} and \cite{KMaximal}).
But without a general asymptotic upper bound, it is
unclear what strings could be considered random for $A_P$, and so we ask:

\begin{que} Is there a suitable notion of a string being random with respect to
  $A_P$? If so, then asymptotically, how many strings are random in this sense?
\end{que}

%% file: computability.tex
\section{Computability of probabilistic complexity}\label{sec:apcomp}

\begin{thm}\thlabel{apcomp}
  For all $\delta\in [0,1)$, the function $w \mapsto A_{P,\delta}(w)$ is
  $\delta$-computable.
\end{thm}

This result is derived as a consequence of the following two theorems, whose
proofs use substantially different methods.
First we examine the uniform computability of $A_{P,\delta}(w)$ as a function of
both $\delta$ and $w$:
\begin{thm}\thlabel{apdeltacomp}
  For any finite alphabet $\Sigma$, the function $(\delta,w) \mapsto
  A_{P,\delta}(w)$ is
  \begin{itemize}
    \item Continuous everywhere on $[0,1) \times \Sigma^\ast$ except on a
      countably infinite set enumerable by a single algorithm;
    \item Computable on $(0,1) \times \Sigma^\ast$ where it is continuous.
  \end{itemize}
  In particular, for every $w$, $A_{P,\delta}(w)$ is computable for all but at
  most $A_D(w) - 2$ many values of $\delta$, and is continuous at $\delta=0$.
\end{thm}

To prove this we need some machinery from computable analysis, and we introduce
the needed background in the next subsection before proceeding to the proof. 
As a function whose input is a Cauchy name for $\delta$, it is currently unknown
whether $A_{P,\delta}(w)$ is computable at $\delta=0$, since the proof of
\thref{apdeltacomp} does not extend to this case. 
If instead one works with a finitary description of $\delta$ via a first-order
formula, then we do have computability at $\delta=0$ and at all discontinuities
of the two-variable function $A_{P,\delta}(w)$ by the next theorem, whose
proof was suggested to the author by Bj{\o}rn Kjos-Hanssen.
The proof of \thref{apcomp} is then completed by observing that $\delta$ is
first-order definable whenever $A_{P,\delta}(w)$ is discontinuous at $(\delta,
w)$, which we show at the end of Section~\ref{sec:compproof}.

\begin{thm}\thlabel{tarski}
  $A_{P,\delta}$ is computable whenever $\delta$ is first-order definable in the
  language $\mathcal{L}$ of real closed fields.
\end{thm}

The precise definition of a real closed field is unnecessary for our purposes,
and may be found, e.g., in \cite[\S3.3]{Marker}. The language $\mathcal{L}$
consists of constant symbols $0$ and $1$ together with a binary relation symbol
$<$ and binary function symbols $+$, $-$, and $\cdot$.
Then $(\R; 0,1,+,-,\cdot,<)$ is a real closed field with these symbols being
given their usual meaning.
Tarski proved that the first-order theory of real closed fields is decidable,
i.e., there is an algorithm which decides if the theory of real closed fields
proves a given $\mathcal{L}$-sentence \cite[Corollary~3.3.16]{Marker}.

\begin{proof}
  By Tarski's theorem, it suffices to show that for a given $k$, $w$, and
  definable number $\delta$, the relation $A_{P,\delta}(w) \leq k$ is equivalent
  to an $\mathcal{L}$-sentence, and that (a G\"odel number for) such a sentence
  can be uniformly computed from $k$ and $w$.

  In what follows, tuples $\bar{a}$ and $\bar{b}$ will always be elements of
  $\R^{k(bk+2)}$ where $\abs{\Sigma}=b$ is fixed. For each $k$, define
  \begin{equation}
    \mathrm{ispfa}_k(\bar{a}) \equiv ~ \bigwedge_{i=1}^{k(bk+2)} (0 \leq a_i
    \leq 1) \land
    \bigwedge_{n=1}^{bk+1} \paren{\sum_{j=nk+1}^{(n+1)k} a_j=1} \land
    \bigwedge_{i=k(bk+1)+1}^{k(bk+2)} (a_i=0 \lor a_i=1).
  \end{equation}
  In words, this means that $\bar{a}$ can be split into $bk+1$ stochastic
  vectors followed by a 0-1 vector, all of length $k$. If $M$ is the PFA defined
  by the tuple $\bar{x}$, write $p_w(\bar{x})$ for $\rho_M(w)$; this is a
  polynomial in the entries of $\bar{x}$ and hence a term in $\mathcal{L}$ which
  can be uniformly computed from $w$.
  Next, given $w$, write $\Sigma^{\abs{w}} \setminus \{w\}$ as $\{ w_1, \dotsc,
  w_n \}$ (with $n$ depending on $w$). Let
  \begin{equation}
    \mathrm{isgap}_{k,w}(g,\bar{a}) \equiv ~ \bigvee_{i=1}^n \paren{
      (p_w(\bar{a}) - p_{w_i}(\bar{a}) = g) \land \bigwedge_{j\neq i} \paren{
    p_w(\bar{a}) - p_{w_j}(\bar{a}) \geq g}}.
  \end{equation}
  Thus $\mathrm{isgap}_{k,w}(g,\bar{a})$ holds if and only if $\gap_M(w) = g$
  where $M$ is the $k$-state PFA defined by $\bar{a}$. It is clear that one can
  uniformly compute the $\mathcal{L}$-formulas $\mathrm{ispfa}_k(\bar{a})$ and
  $\mathrm{isgap}_{k,w}(g,\bar{a})$ from $k$ and $w$.
  If $\delta$ is defined by the formula $\varphi(x)$, then we have
  $A_{P,\delta}(w) \leq k$ if and only if
  \begin{equation}\label{eq:apdformula}
    \exists \bar{a} \exists g_1 \exists g_2 ~ \paren{\mathrm{ispfa}_k(\bar{a}) \land
    \mathrm{isgap}_{k,w}(g_1,\bar{a}) \land \varphi(g_2) \land (g_1 > g_2)},
  \end{equation}
  and $A_P(w) \leq k$ if and only if this sentence holds with $g_2$ replaced by
  $0$.
\end{proof}

As we said above, the proof of \thref{apcomp} will be completed at the end of
Section~\ref{sec:compproof}, after \thref{apdeltacomp} has been proven. We now
go on to review the needed background for the latter proof.


\subsection{Background in computable analysis}

We follow the approach of \cite{DowneyMelnikov}. For a separable metric
space $(X,d)$, suppose we are given an enumeration $\alpha \colon \N\to X$ of a
dense subset of $X$. Fix some enumeration $(q_i)_{i \in \N}$ of $\Q$. Then we
say $X$ is a \bdef{computable metric space} if $d\colon X\times X\to\R$ is
computable when restricted to the range of $\alpha$, in the sense that the set
\[
  \set{(i,j,n,m)\in \N^4 \stc q_i < d(\alpha(n),\alpha(m)) < q_j}
\]
is c.e. The function $\alpha$ gives rise to a canonical computable enumeration
of a basis for the topology on $X$, namely
\[
  \ang{i,j} \mapsto B_{q_j}(\alpha(i)),
\]
where $B_q(x)$ is the open ball of radius $q$ centered at $x\in X$. We will from
now on refer to the sets in this canonical enumeration as \bdef{basic open
balls}. 
We may refer to a procedure as ``outputting an open ball'' or ``listing
open balls'' when we really mean that it produces an index $\ang{i,j}$ for a
basic open ball, or a list of such indices.

A \bdef{name} for a point $x\in X$ is a list $N^X_x$ (in any order) of all basic
open balls in $X$ containing $x$. If $(X,d_X)$ and $(Y,d_Y)$ are two computable
metric spaces, a function $f\colon X\to Y$ is \bdef{computable} if there is a
Turing functional which sends $N^X_x$ to $N^Y_{f(x)}$ for all $x\in X$. A
\bdef{Cauchy name} for a point $x$ is a sequence $(x_n)\subset D$ converging to
$x$ such that for all $n$, $d(x_n,x_{n+1}) < 2^{-n}$. One can compute a Cauchy
name for $x$ from $N^X_x$ by first finding a subsequence of basic open balls
listed in $N^X_x$ with exponentially decreasing radii, then taking their
centers. Conversely, one can compute a name $N^X_x$ from a Cauchy name: if
$(x_n)$ is a Cauchy name for $x$ and $B_q(y)$ is any basic open ball, then
$d(x,y)<q$ iff $d(x_n,y) < q-2^{-n}$ for some $n$, and the latter will be
witnessed in finite time since by assumption $d(x_n,y)$ is computable in the
sense given above. Neither algorithm depends on $x$, and so if $f$ is computable
in the above sense, then there is also a uniform computable procedure mapping a
Cauchy name for $x$ to a Cauchy name for $f(x)$ for all $x$. Every computable
function is continuous.

The real line $\R$ is a computable metric space with the usual Euclidean metric,
taking $D=\Q$. A computable real number is a number having a computable Cauchy
name, viewed as an element of Baire space. If $f,g\colon X\to\R$ are computable
functions, then so are $f+g$, $f-g$, $fg$, $\max\{f,g\}$, and $\min\{f,g\}$. In
particular, by taking both $f$ and $g$ to be the identity map on $\R$, we get
that the function $(x,y)\mapsto \max\{x,y\}$ is computable. If given $x\neq y$,
one can also decide in finite time from their Cauchy names which is larger. 

A computable metric space $X$ is \bdef{computably compact} if there is a
computable function which outputs a finite open cover of $X$ by basic open balls
of radius at most $2^{-n}$, given input $n$. If $f\colon X\to\R$ is computable
and $X$ is computably compact, then $\sup_{x\in X} f(x)$ and $\inf_{x\in X}
f(x)$ are computable numbers, and this is uniform in $f$ (identifying $f$ with
an index for an oracle Turing machine mapping $x\mapsto f(x)$).


\subsection{Proofs of \thref{apdeltacomp} and \thref{apcomp}}\label{sec:compproof}
For any $k\geq2$, let $\A_k$ denote the space of $k$-state PFAs over a fixed
finite alphabet $\Sigma$, where we identify $\Sigma$ with $\{0, \dotsc, b-1\}$.
To be precise, define
\begin{align*}
  \A_k = \big\{ & (\vec{\pi}, P_0, P_1, \dotsc, P_{b-1}, \vec{\eta}) :
    \text{$\vec{\pi} \in [0,1]^k$ is a probability vector,}\\
    &\text{each $P_\sigma$ is a $k\times k$ stochastic matrix, and $\vec{\eta}
  \in \{0,1\}^k$} \big\} \subset [0,1]^{2k+bk^2}.
\end{align*}

If $A\in \A_k$, write the components of $A$ as $\vec{\pi}^A, P_0^A, \dotsc,
P_{b-1}^A$, and $\vec{\eta}^A$. Also write $M^A$ for the vector $(\vec{\pi}^A,
P_0^A, \dotsc, P_{b-1}^A)$. We give $\A_k$ the uniform
(maximum) distance $d(\cdot,\cdot)$, i.e., that induced from the product
topology on $[0,1]^{2k+bk^2}$. (The euclidean distance would work just as well.)
Then $\A_k$ is a computably compact metric space. There are several easy ways to
see this, but we give a direct proof for convenience. Let $Q_k$ be the set of
rational $k$-state PFAs, that is, the set of $A\in \A_k$ such that all entries
of $M^A$ are rational, given as quotients of natural numbers. Clearly $Q_k$ has
a computable enumeration and is dense in $\A_k$, and $d(A,B)$ is computable for
any $A,B\in Q_k$, hence $\A_k$ is a computable metric space. Then for any fixed
$n$, one can enumerate all $A\in Q_k$ such that every entry of $M^A$ is equal to
$j2^{-n-1}$ for some $j\in \{0, \dotsc, 2^{n+1} \}$. The set of $B_{2^{-n}}(A)$
for all such $A$ is a finite open cover of $\A_k$ by basic open balls of radius
at most $2^{-n}$, so $\A_k$ is computably compact by definition.

The function $(A,w)\mapsto \rho_A(w)$ is computable, because it is a polynomial
in the entries of $A$ resulting from multiplication of $\vec{\pi}^A$,
$\vec{\eta}^A$, and the matrices $P_\sigma^A$ in an order determined by $w$.
Therefore $(A,w) \mapsto \gap_A(w)$ is the minimum of finitely many computable
functions and hence itself computable, as is $A\mapsto \gap_A(w)$ for any fixed
$w$.
Now let
\begin{equation}
  \gamma^k(w) = \max_{A\in \A_k} \gap_A(w).
\end{equation}
For each $k$ and $w$, $\gamma^k(w)$ is a computable real number, because it is
equal to the supremum of the computable function $A\mapsto \gap_A(w)$ over the
computably compact space $\A_k$. And since the procedure to compute $\gap_A(w)$
is uniform in $w$, the function $(k,w) \mapsto \gamma^k(w)$ is computable. 
Finally, let
\begin{equation}\label{eq:Edef}
  E = \set{ (\gamma^k(w),w) \stc 2\leq k \leq A_D(w)-1, ~ w\in \Sigma^\ast,
  0 < \gamma^k(w) < 1} \subset (0,1)\times \Sigma^\ast.
\end{equation}
This will turn out to be exactly the set of discontinuities of
$A_{P,\delta}(w)$, and it can clearly be enumerated by a single algorithm by
definition.
\thref{upperbound}(ii) implies that $A_{P,\delta}(w)$ is continuous at
$(0,w)$ for all $w$. Continuity on the remainder of the complement of $E$ will
follow from the computability argument below.

That $E$ is countably infinite is a consequence of the following fact of
potential independent interest, whose proof establishes that in some sense, a
2-state PFA giving a gap of 1 to even a single word (with more than three
letters) behaves much like a DFA as far as $A_P$ is concerned.

\begin{lem} For any $w$ with $\abs{w} \geq 4$, $\gamma^2(w)=1$ iff $w$ is
  constant.
\end{lem}
\begin{proof} The right-to-left implication is immediate, since then $A_D(w) =
  2$. For the other direction, assume for sake of contradiction that $w$ is
  nonconstant, and that $\gamma^2(w)=1$ is witnessed by the IFS with starting
  value $x_0$ and maps $f_jx = a_j + b_jx$ for each letter $j \in \Sigma$. Then
  $\rho(w)=1$ and $\rho(y) = 0$ for every other $y$ of length $\abs{w}$, and if
  $w = z^\smallfrown i$ where $i\in \Sigma$, then in particular $f_i
  \rho(z) = 1$ and $f_j \rho(z) = 0$ for all $j\neq i$. Now, if the
  range of $f_i$ omits the value $0$, then $\rho(i^n)>0$ for all $n$, regardless
  of the value of $x_0$. Then either $w$ is constant or $\gap(w)<1$, a
  contradiction, and we may thus assume the range of $f_i$ includes both $0$ and
  $1$. By drawing a picture, one sees that only $f_ix=x$ and $f_ix= 1-x$ are
  possible. If $f_ix$ is the identity then only constant strings may be
  witnessed, so we can assume that $f_ix = 1-x$.

  If $f_ix = 1-x$, then $\rho(z)=f_i\inv(1) = 0$, so $f_j0=0$ and thus
  $f_jx = b_jx$ for all $j \neq i$. We can take $b_j < 1$, as otherwise $f_j$ is
  the identity map and only constant strings can be witnessed. If $b_j=0$ for
  some $j$, so that $f_j \equiv 0$, then every string ending in $ji$ has
  probability $1$, thus maximal probabilities are nonunique starting at length
  3, a contradiction. Then $0<b_j<1$ for all $j\neq i$, and this means once an
  orbit leaves $\{0,1\}$ it can never return to either value.
  In particular, $x_0\in \{0,1\}$. If $x_0=0$, then for all $n\geq1$ we
  have $\rho(ji^{ 2n-1}) = \rho((ji)^{2n}) = 1$ among even-length strings and
  $\rho(i^{2n+1}) = \rho(j^2 i^{2n-1}) = 1$ among odd-length strings.
  If $x_0 = 1$, then for all $n\geq 1$ we have $\rho(i^{2n}) = \rho(ij^{2n-2}i)
  = 1$ among even-length strings and $\rho(ij^{2n-1}i) = \rho(iji^{2n-1}) = 1$
  among odd-length strings. Either way, uniqueness of maxima is lost starting at
  length at most 4, so $\gap(w) < 1$ and by contradiction the proof is complete.
\end{proof}

There are infinitely many nonconstant $w$ with $\abs{w} \geq 4$ and $A_P(w)=2$,
of course, by \thref{classification2-full}. For such $w$, the lemma implies that
$0 < \gamma^2(w) < 1$, so that $(\gamma^2(w), w) \in E$ and in particular $E$ is
infinite.

We now show that $A_{P,\delta}(w)$ is discontinuous on $E$ and computable on
the complement of $E$, minus the points with $\delta=0$.
Endow $\Sigma^\ast$ with the discrete topology in its standard metrization,
i.e., $d(x,y)=1$ iff $x\neq y$. Then we give $[0,1)\times \Sigma^\ast$ the
product metric, that is, $d\paren{ (\alpha,x), (\beta,y) } = \max\{
\abs{\alpha-\beta}, d_{\Sigma^\ast}(x,y)\}$. The codomain $\N$ of
$A_{P,\delta}(w)$ also has the discrete topology as a subset of $\R$. Now,
$A_{P,\delta}(w)$ is continuous at $(\delta,w)$ iff for all $\varepsilon >0$
there is an $\eta>0$ such that $d\paren{ (\delta,w) - (\delta',w')} <\eta$
implies $\abs{A_{P,\delta'}(w) - A_{P,\delta}(w)} < \varepsilon$---so that
actually $\abs{\delta-\delta'}<\eta$ implies $A_{P,\delta'}(w) = A_{P,
\delta}(w)$ (since $\Sigma^\ast$ and $\N$ both have the discrete topology). If
$\delta=\gamma^k(w)$ for some $k$ and $w$, then by definition of $\gamma^k$
there is no $\delta'<\delta$ such that $A_{P, \delta'}(w) = A_{P, \delta}(w)$,
because there is a $k$-state PFA having a gap greater than $\delta'$ for $w$ but
not one having a gap greater than $\delta$. Hence $A_{P,\delta}(w)$ is
discontinuous at every point of $E$.

Finally, let $(\delta,w)\notin E$ be given with $\delta \neq 0$. Under these
hypotheses, for any $k\geq 2$, we have $\delta> \gamma^k(w)$ if and only if
$A_{P,\delta}(w) >k$, because in this case there is no $A\in \A_k$ exhibiting
the required gap. Conversely, $\delta < \gamma^k(w)$ if and only if
$A_{P,\delta}(w) \leq k$. To compute $A_{P,\delta}(w)$, then, decide for each
$k=2,3,\dotsc,A_D(w)$ whether $\delta$ or $\gamma^k(w)$ is greater.
The least $k$ such that $\delta < \gamma^k(w)$ is exactly equal to
$A_{P,\delta}(w)$. 
It is clear that this procedure does not depend on $\delta$ or $w$, and the
proof of \thref{apdeltacomp} is complete. \QED

We are now ready to finish the proof of \thref{apcomp} by showing that $\delta$
is definable in the language $\mathcal{L}$ whenever $(\delta,w) \in E$, where
$\mathcal{L}$ was defined after the statement of \thref{tarski} and $E$ is given
by \eqref{eq:Edef}, so that \thref{tarski} applies to $\delta$. Let
\begin{equation}
  \begin{aligned}
  \mathrm{ismaxgap}_{k,w}(g) \equiv &\paren{ \exists \bar{a} \br{
  \mathrm{ispfa}_k(\bar{a}) \land \mathrm{isgap}_{k,w}(g,\bar{a}) }}\\
  &\land \paren{ \forall \bar{b} \br{ \mathrm{ispfa}_k(\bar{b}) \rightarrow
  \bigvee_{i=1}^n \paren{ p_w(\bar{b}) - p_{w_i}(\bar{b}) \leq g}}},
\end{aligned}\end{equation}
where all notation is as in the proof of \thref{tarski}. In words, this says
there is a $k$-state PFA giving $w$ gap $g$ and that all $k$-state PFAs give $w$
gap at most $g$. Hence $\mathrm{ismaxgap}_{k,w}(g)$ holds iff $g = \gamma^k(w)$.
Since $\mathrm{ismaxgap}_{k,w}$ can evidently be uniformly computed from $k$ and
$w$, we are done. \QED


\subsection{Remarks}
The proof of \thref{tarski} can be easily adapted to show that many variants on
$A_P$ and $A_{P,\delta}$ are computable. For example, one might change
$A_{P,\delta}$ to $A_{P, \geq \delta}$ by requiring a witness $M$ to satisfy
$\gap_M(w) \geq \delta$ rather than $>\delta$, and then replace $g_1>g_2$ in
\eqref{eq:apdformula} by $g_1\geq g_2$ to show $A_{P, \geq\delta}$ is
computable. Something similar can be done for the variants discussed in
Section~\ref{sec:relax} below.

The function $(k,w) \mapsto \gamma^k(w)$ may be of interest in and of
itself, as briefly mentioned in the following section.
We have $0 \leq \gamma^2(w) \leq \gamma^3(w) \leq \cdots \leq \gamma^{A_D(w)}(w)
= 1$, with $\gamma^k(w) = 0$ if and only if $k<A_P(w)$. The number
$\gamma^k(w)$ is never negative since one can always make a PFA accepting every
word with the same probability by setting all transition matrices to the
identity matrix. Furthermore, \thref{apxy} implies $\gamma^k(z) \geq
\gamma^k(wz)$ for all $w$, $z$, and $k$. This comes close to justifying the
empirical observation made in Section~\ref{sec:firstresults} that gaps tend to
decrease for longer words. A result to the effect that $\gamma^k(w) \geq
\gamma^k(wz)$ would put the observation on fully rigorous ground.

%% file: otherdefs.tex
\section{Other approaches to probabilistic complexity}\label{sec:otherdefs}

\subsection{Relaxing the definition of a PFA}\label{sec:relax}
We saw earlier that $A_P$ shares the property of $A_D$ that the complexity of a
string is not necessarily equal to that of its reversal. In addition, there are
strings whose PFA complexity is known to be witnessed by a PFA with dead states.
One might try to solve these problems by relaxing the definition of a PFA to
directly generalize an NFA (rather than a DFA). 
NFAs are allowed to have rows of all zeros in their transition matrices, and
also have the property that different out-transitions from the same state and
for the same letter are not weighted differently---they are simply all possible.
The same applies to the initial state distribution.

To directly translate these properties to a generalization of a PFA, one would
need to require that all nonzero entries of $\vec{\pi}$ are equal, as well as
the same for $\vec{\eta}$, and that all
nonzero entries of all the matrices $P_\sigma$ are equal to the same number
(which may result in the row sums being different). One can see that the proof
of the first part of \thref{reversal} can be recovered for the class
$\mathscr{N}$ of such automata. 
Since $\mathscr{N}$ is closed under switching $\vec{\pi}$ and $\vec{\eta}$ and
taking the transpose of every transition matrix, if $\tilde{A}_P$ is the
corresponding complexity notion where $\tilde{A}_P(x)$ must be witnessed by a
member of $\mathscr{N}$, then one has $\tilde{A}_P(\rev{x}) = \tilde{A}_P(x)$
for all $x$.
(Moreover $\tilde{A}_P(x) \leq A_N(x)$.)
However, this is not a very natural class of automata to consider and it is
certainly not a generalization of a PFA.

Instead of trying to design a specific class of automata in an attempt to
recover properties of $A_N$, it might make more sense to define a unified
complexity notion which takes as parameter a family of automata and study its
properties in general. In \cite{Tur69}, Turakainen introduced \bdef{generalized
(probabilistic) finite automata} (GPFAs), which are finite-state automata whose
operation is described as follows:
\begin{itemize}
  \item The initial state of the machine is an arbitrary real row vector.  
  \item Transitions between states are described by multiplication of arbitrary
    real square matrices.
  \item The final state of the machine is again an arbitrary real column vector.
\end{itemize}
So GPFAs are like PFAs except that the entries of $\vec{\pi}$, $\vec{\eta}$, and
each $P_\sigma$ can be any real numbers. Turakainen proved the remarkable
fact that GPFAs have in a sense the same descriptive power as PFAs: if one
also allows a cut-point in the context of a GPFA to be any real number, then the
class of languages accepted by GPFAs is exactly the class of stochastic
languages.

This suggests that it is not too unreasonable to throw the gates open and
consider a version of $A_P$ that allows any GPFA. Let $\mathscr{G}$ be the set of
all GPFAs. For any family $\mathscr{F}\subseteq \mathscr{G}$, let $\mathscr{F}_k$
be the set of members of $\mathscr{F}$ having $k$ states. Then define the
\bdef{$\mathscr{F}$-automatic complexity} of a word $x\in \Sigma^\ast$ to be
\[
  A_{\mathscr{F}}(x) = \min \set{k\stc \exists F\in \mathscr{F}_k 
    \text{ such that } \gap_F(x) > 0}.
\]

For example, $A_P$ as defined before coincides with $A_{\A}$ if
$\A \subset \mathscr{G}$ is the set of all PFAs. One can also define
$A_{\mathscr{F},\delta}$ for any $\delta \geq 0$ by analogy with $A_{P,\delta}$.
We have that for all $x$,
\[
  A_{\mathscr{E}}(x) \leq A_{\mathscr{F}}(x) \quad\text{whenever}\quad
  \mathscr{E} \supseteq \mathscr{F},
\]
so that in particular $A_{\mathscr{G}}(x) \leq A_{\mathscr{F}}(x)$ for every
$\mathscr{F}$ and $x$. We have not investigated $A_{\mathscr{F}}$ in general,
and it is unclear how coarse of a measurement it might be. As a motivating
question, we could ask
\begin{que} Is $A_{\mathscr{G}}(x)\leq 2$ for all binary strings $x$?
\end{que}

It is at least true that $A_{\mathscr{G}}(x) = A_{\mathscr{G}}(\rev{x})$ for all
$x$, by the same observation we made above for $\tilde{A}_P$---and indeed
$A_{\mathscr{G},\delta}(x) = A_{\mathscr{G},\delta}(\rev{x})$ for all $x$ and
$\delta \geq 0$, 
since one can simply switch $\vec{\pi}$ and $\vec{\eta}$ and replace
$P_\sigma$ with $P_\sigma^T$ for all $\sigma$ to change a witness for
$A_{\mathscr{G},\delta}(x)$ into one for $A_{\mathscr{G},\delta}(\rev{x})$
while preserving the acceptance probability of every word.
Together with \thref{apxy}, whose proof goes through
verbatim for $A_{\mathscr{G},\delta}$, this implies that $A_{\mathscr{G},
\delta}(xyz) \geq A_{\mathscr{G}, \delta}(y)$ for all $x$, $y$, and $z$, like
$A_D$ and $A_N$. \thref{tarski} also holds for $A_\mathscr{G}$ simply by making
$\mathrm{ispfa}_k(\bar{a})$ true for all $\bar{a}$ and all $k$.

A potentially helpful observation here is that the ability to have unbounded
real entries does not really confer any advantage as far as the complexity of
individual strings is concerned. For any GPFA $M$, if $C$ is the largest
absolute value of any entry of $\vec{\pi}^M$, $\vec{\eta}^M$, and the matrices
$P_\sigma^M$, then one could divide all these matrices and vectors by $C$ to
obtain a GPFA $M'$ with entries in $[-1,1]$ such that
\[
  \rho_M(x) < \rho_M(y) \iff \rho_{M'}(x) < \rho_{M'}(y)
\]
whenever $\abs{x}=\abs{y}$. Hence if $\mathscr{S}$ is the set of GPFAs whose
entries are all in $[-1,1]$, we have $A_{\mathscr{S}}(x) = A_{\mathscr{G}}(x)$
for all $x$. In addition, the direct analogue of \thref{apdeltacomp} holds for
$A_{\mathscr{S},\delta}$, because $\mathscr{S}_k$ is now a computably
compact metric space for each $k$, unlike $\mathscr{G}_k$.

One advantage of $A_P$ that appears to be immediately lost in passing to
$A_{\mathscr{G}}$ or $A_{\mathscr{S}}$ is the dimension reduction of the IFS
approach, and the dynamical analysis made more tractable by it. Since the
correspondence between PFAs and IFSs relies explicitly on the transition
matrices being stochastic, perhaps one could allow only generalized stochastic
transition matrices, with any real entries permitted as long as each row sums to
1. This notion would for example allow us to describe $0100$ in two states via
\[
  P_0 = \begin{pmatrix} -1 & 2\\ 1/2 & 1/2\end{pmatrix}, \quad
  P_1 = \begin{pmatrix} 1/2 & 1/2\\ 1 & 0\end{pmatrix}, \quad
  \vec{\pi} = (0,1), \quad \vec{\eta} = \begin{pmatrix} 1\\0\end{pmatrix},
\]
whereas $A_P(0100)=3$, so strictly greater compression is achieved. This
automaton is equivalent to the IFS with $f_0(x) = \frac{1}{2} - \frac{3}{2}x$,
$f_1(x) = 1-\frac{1}{2}x$, and $x_0=0$. (Other strings with $A_P=3$ which have
complexity $2$ according to this notion include $01000$, $01011$, and $01100$.)
Unfortunately, uniformly rescaling the transition matrices as with $\mathscr{S}$
no longer works here, so the set of allowed transition probabilities is
unbounded and we lose uniform computability of the analogue of $A_{P,\delta}$,
i.e., the proof of \thref{apdeltacomp} cannot be recovered. Of course, the proof
of \thref{tarski} can still be easily modified to fit this case by simply
dropping the requirement that probabilities lie between $0$ and $1$ from the
formula $\mathrm{ispfa}_k$.

\subsection{Gap structure function}
We saw in the proof of \thref{apdeltacomp} that the function $\gamma^k(w)$ mapping
$w$ to the maximum value of $\gap_M(w)$ among all $k$-state $M$ is computable.
It could be interesting to study $\gamma^k(w)$ as a parametrized complexity
measure in itself. Intuitively, $w\mapsto \gamma^k(w)$ measures how well $w$ is
described by the model class of $k$-state PFAs---the widest margin of
probability by which $w$ can be recognized by any such PFA. This relates
$\gamma^k(w)$ at least philosophically to the Kolmogorov structure function,
which measures the minimal size of a set of strings containing $w$ which can be
described by a Turing machine of size at most $k$, and hence captures in a sense
how well $w$ can be singled out by such machines. Similar functions have also
been considered by Kjos-Hanssen \cite{K20}, who introduced both a structure
function and a dual structure function for the NFA complexity. His dual
structure function is in part motivated by having a simple domain and
complicated range, rather than the other way around as for his regular structure
function for $A_N$. 
This is even more true for $\gamma^k(w)$ in contrast with its dual $A_{P,\delta}
(w)$, at least if one is interested in computability: it maps a string and
natural number to a Cauchy name for a real number, rather than mapping a string
and some representation of a real (either a Cauchy name or a first-order
formula, depending) to a natural number.

\subsection{Least number of bits of a witness}
Heuristically it appears that witnesses for the PFA complexity of many strings
are relatively complicated; this certainly seems to be the case for most strings
with $A_P=2$, as pointed out below.
If one is interested solely in compression, it might make the most sense to
measure the complexity of $w$ as the least number of bits required to describe
an $M$ having $\gap_M(w)>0$, or perhaps $\gap_M(w) > \delta$ for a parameter
$\delta$. One potential drawback of this approach is that it is not obvious
whether this measure is computable, although this depends on the precise
definition used. The least number of bits also depends on the choice of
encoding, and so this measure would only be defined up to an additive constant,
like the Kolmogorov complexity. Not only that, but it could well be that the
simplest PFAs achieving a positive gap are very often DFAs, and in that case one
could argue it is hardly a satisfying notion of PFA complexity.

\subsection{Measure of the set of witnesses}
We conclude by mentioning one more idea for modifying $A_P$ and $A_{P,\delta}$,
with the aim of refining the numerical measurement itself. In the proof of for
example \thref{rev1}, we saw that although all strings $0^n1^m$ have complexity
2, as $n$ increases, $x_0$ must be chosen in a narrower and narrower range in order
for the IFS to witness $0^n1^m$. The coefficient $b$ must also be made
arbitrarily close (but not equal) to 1. Something similar is true of the other
subcases of the proof of \thref{classification2-reverse}. Thus it is in a sense
more complicated to witness the complexity of a string the longer its prefix is.
So, we could introduce a real-valued complexity measure that accounts for that
difference as follows. Let $\mu$ be a Borel probability measure with full
support on $\A_k$, the space of $k$-state PFAs. Let $G^k(x) =
\gap_{\bullet}(x)\inv( (0,1])$ be the set of $k$-state witnesses for $A_P(x)\leq
k$, and let
\[
  A_\mu(x) = A_P(x)+1-\mu(G^k(x)).
\]
We can also let $G^k_\delta(x) = \gap_{\bullet}(x)\inv( (\delta,1])$ and define
\[
  A_{\mu,\delta}(x) = A_{P,\delta}(x) + 1 - \mu(G^k_\delta(x)).
\]

Since $\gap$ is a computable function on a computably compact metric space,
$G^k(x)$ and $G^k_\delta(x)$ are c.e.\ open, meaning the indices of all basic
open balls contained in each of them can be computably enumerated. 
In particular, all these sets have positive $\mu$-measure if nonempty.
Thus $A_\mu(x)$ assigns $x$ a value strictly
between $A_P(x)$ and $A_P(x)+1$, and $A_{\mu,\delta}(x)$ is strictly between
$A_{P,\delta}(x)$ and $A_{P,\delta}(x)+1$. Moreover, in at least the case of
binary strings with $A_P(x)=2$, if $x$ is a string such as $0^n101$ which can only be
witnessed by a PFA that also witnesses $0^n1(01)^m$ for all $m$, then those
strings receive exactly the same value of $A_\mu$ as $x$ does. This makes sense,
because these extensions of $x$ in a sense do not require any further effort to
find a witness. The latter observation holds for any measure $\mu$. The goal in
defining $A_\mu$ (or $A_{\mu,\delta}$) would then be to find a suitable $\mu$
which gives sets like $G^k(x)$ (or $G^k_\delta(x)$) large measure for strings
like $x=(01)^n$ which are easy to witness, while giving smaller measure to
$G^k(x)$ (or $G^k_\delta(x)$) for strings whose witnessing automata require a
more precise configuration.

A natural choice would be for $\mu$ to be induced by Lebesgue measure on the
unit simplex in $\R^{k-1}$, identifying $k$-state PFAs with $(k-1)$-dimensional
affine IFSs as in Section~\ref{sec:ifs}. If $\abs{\Sigma} =b$, one could take
the $(bk+1)$-fold product of the $(k-1)$-dimensional Lebesgue measure with itself,
one for each stochastic row vector in an element of $\A_k$, and average it over
the $2^k-2$ connected components of $\A_k$ corresponding to nontrivial choices
of $\vec{\eta}$. The result is a fully supported computable probability measure
$\mu$ on $\A_k$.

\begin{que}
  Does this $\mu$ lead to satisfactory, and in particular computable, complexity
  measures $A_\mu$ and $A_{\mu,\delta}$?
\end{que}

\begin{que} How should the definitions of $A_\mu$ and $A_{\mu,\delta}$ account
  for the fact that lower-complexity strings are also witnessed by members of
  $\A_k$? How should one deal with the likely problem of the sets $G^k(x)$
  generally having high measure when $A_P(x)<k$, which would make $A_\mu$
  clustered near $k+1$ among strings having $A_P(x)=k$? 
\end{que}